\documentclass[prd,aps,twocolumn,nofootinbib,superscriptaddress,showpacs]{revtex4-1}
\usepackage{amsfonts}
\usepackage{amsmath}
\usepackage{amssymb}
\usepackage{bm}
\usepackage{dcolumn}
\usepackage[dvips]{graphicx}
\usepackage{graphics}
\usepackage[latin1]{inputenc}
\usepackage{latexsym}
\usepackage{rotating}
\usepackage[colorlinks=true]{hyperref}
\usepackage{xspace} 
\usepackage[usenames]{color}
\usepackage{mathrsfs}
\usepackage{multirow}
\usepackage{pifont}
\usepackage{enumitem}

\definecolor {darkgreen}{rgb}{0.2,0.7,0.2}
\definecolor{purple}{rgb}{0.5,0,0.5}

\newcommand\be{\begin{equation}}
\newcommand\ba{\begin{eqnarray}}
\newcommand\ee{\end{equation}}
\newcommand\ea{\end{eqnarray}}

\newcommand\bw{\begin{widetext}}
\newcommand\ew{\end{widetext}}

\newcommand{\nn}{\nonumber}

\newcommand{\GW}{{\mbox{\tiny GW}}}

\newcommand{\GR}{{\mbox{\tiny GR}}}

\newcommand{\MAT}{{\mbox{\tiny mat}}}

\newcommand{\mrm}{\mathrm}

\begin{document}

\title{Probing Gravitational Parity Violation with Gravitational Waves \\ from Stellar-mass Black Hole Binaries}

\author{Kent Yagi}
\affiliation{Department of Physics, University of Virginia, Charlottesville, Virginia 22904, USA}
\affiliation{Department of Physics, Princeton University, Princeton, New Jersey 08544, USA}

\author{Huan Yang}
\affiliation{Perimeter Institute for Theoretical Physics, Waterloo, Ontario N2L 2Y5, Canada}
\affiliation{University of Guelph, Guelph, Ontario, N2L 3G1, Canada}
\affiliation{Department of Physics, Princeton University, Princeton, New Jersey 08544, USA}
\date{\today}

\begin{abstract} 

The recent discovery of gravitational wave events has offered us unique testbeds of gravity in the strong and dynamical field regime. 
One possible modification to General Relativity is the gravitational parity violation that arises naturally from quantum gravity.
Such parity violation gives rise to the so-called ``amplitude birefringence" in gravitational waves, where one of the circularly-polarized mode is amplified while the other one is suppressed during their propagation.
In this paper, we study how well one can measure gravitational parity violation via the amplitude birefringence effect of gravitational waves sourced by stellar-mass black hole binaries.
We choose Chern-Simons gravity as an example and work within an effective field theory formalism to ensure that the approximate theory is well-posed.
We consider gravitational waves from both individual sources and stochastic gravitational wave backgrounds.
Regarding bounds from individual sources, we estimate such bounds using a Fisher analysis and carry out Monte Carlo simulations by randomly distributing sources over their sky location and binary orientation.
We find that the bounds on the scalar field evolution in Chern-Simons gravity from the recently-discovered gravitational wave events are too weak to satisfy the weak Chern-Simons approximation, while aLIGO with its design sensitivity can place meaningful bounds.
Regarding bounds from stochastic gravitational wave backgrounds, we set the threshold signal-to-noise ratio for detection of the parity-violation mode as 5 and estimate projected bounds with future detectors assuming that signals are consistent with no parity violation.
In an ideal situation where all the source parameters and binary black hole merger rate history is known \emph{a priori}, we find that a network of two third-generation detectors is able to place bounds that are comparable or slightly stronger than binary pulsar bounds. 
In a more realistic situation where one does not have such information beforehand, approximate bounds can be derived if the regular parity-insensitive mode is detected and the peak redshift of the merger rate history is known theoretically.
Since gravitational wave observations probe either the difference in parity violation between the source and the detector (with individual sources) or the line-of-sight cosmological integration of the scalar field (with gravitational wave backgrounds), such bounds are complementary to local measurements from solar system experiments and binary pulsar observations.

\end{abstract}

\maketitle

\section{Introduction}  

\subsection{Background}

The historic detection of GWs from stellar-mass black hole (BH) and neutron star binaries~\cite{Abbott:2016blz,Abbott:2016nmj,TheLIGOScientific:2016pea,Abbott:2017vtc,Abbott:2017oio,TheLIGOScientific:2017qsa,Abbott:2017gyy} has marked the birth of gravitational-wave (GW) astronomy. Such discoveries also opened new avenues for testing gravity~\cite{TheLIGOScientific:2016src,Yunes:2016jcc,Abbott:2017vtc,Abbott:2017oio,Monitor:2017mdv}. These new sources have offered us unique testbeds for probing gravity in the strong and dynamical field regime, which was previously inaccessible with solar system experiments~\cite{TEGP,Will:2014kxa}, binary pulsar~\cite{stairs,Wex:2014nva} and cosmological~\cite{Jain:2010ka,Clifton:2011jh,Joyce:2014kja,Koyama:2015vza,Salvatelli:2016mgy} observations. The LIGO Scientific Collaboration and Virgo Collaboration (LVC) carried out several different tests of General Relativity (GR) with the events that they discovered~\cite{TheLIGOScientific:2016src,Abbott:2017vtc,Abbott:2017oio,Monitor:2017mdv} while various fundamental aspects of GR have been tested in~\cite{Yunes:2016jcc}\footnote{These GW sources were also used to explore the nature of BH spacetime~\cite{Chirenti:2016hzd,Konoplya:2016pmh,Abedi:2016hgu,Ashton:2016xff,Abedi:2017isz,Brustein:2017kcj,Brustein:2017koc}.}. 
References~\cite{TheLIGOScientific:2016src,Yunes:2016jcc} focused on constraining non-GR modifications in the GW phase.
Thus, one important aspects of gravity not considered in these previous works is the effect of gravitational parity violation on GW propagation, which affects the GW amplitude. 

As an example, gravitational parity violation is present in Chern-Simons (CS) gravity~\cite{jackiw,CSreview}\footnote{Other ways to break gravitational parity includes a model in which the right- and left-handed GWs couple to different gravitational constants~\cite{Contaldi:2008yz}.}, in which a scalar field is linearly coupled to the Pontryagin density at the level of the action. Such a theory can be motivated from a chiral anomaly in the standard model~\cite{Weinberg:1996kr}, gravitational anomaly cancelation in heterotic superstring theory~\cite{polchinski2,Green:1987mn}, loop quantum gravity~\cite{Ashtekar:1988sw,alexandergates,Taveras:2008yf,calcagni,Gates:2009pt} and effective field theories for inflation~\cite{Weinberg:2008hq}. Since the theory contains third derivatives in the field equations and is likely to be ill-posed~\cite{Delsate:2014hba}, one needs to treat it as an effective field theory and work within the small CS approximation (ensuring that the CS corrections are much smaller than GR terms)\footnote{See~\cite{Cayuso:2017iqc} for an alternative approach to cure pathologies in the theory.}. In terms of such an effective field theory approach, the Pontryagin term in the action corresponds to the parity violating term with the lowest mass dimension. 

Bounds on the local evolution of the scalar field in CS gravity have been studied in some literature. Solar system bounds were derived in~\cite{Smith:2007jm} from the frame-dragging measurement by LAGEOS. Binary pulsar bounds were originally found in~\cite{Yunes:2008ua} and was later corrected in~\cite{AliHaimoud:2011bk}, which turned out to be stronger than the solar system bound by more than three orders of magnitude. One can also use the vacuum instability in CS gravity to place bounds on the scalar field evolution that depends on the cutoff momentum scale for effective field theory~\cite{Dyda:2012rj}. A quantum interferometry can also place a bound that is comparable to the solar system bound~\cite{Okawara:2012bi} (see also~\cite{Okawara:2013wc,Kikuchi:2014mva}).

Parity violation has an interesting effect on the GW propagation called \emph{amplitude birefringence}~\cite{jackiw,Alexander:2004wk}, by which the right- (left-) handed GWs are enhanced/suppressed (suppressed/enhanced) as they propagate over a cosmological distance.
GW amplitude birefringence for astrophysical sources in CS gravity was previously studied in~\cite{Alexander:2007kv,Yunes:2008bu} for space-based interferometers and in~\cite{Yunes:2010yf} for coincident GW and gamma-ray observations.  

With the presence of parity-breaking mechanisms, GWs are modified not only during their propagation, but also from their generation. For example, gravitational and scalar radiation were calculated in~\cite{pani-DCS-EMRI} for extreme-mass-ratio inspirals (EMRIs) using the BH perturbation method, while those for comparable mass binary inspirals were calculated in~\cite{Yagi:2011xp} within the post-Newtonian (PN) approximation. Gravitational waveforms for EMRIs were derived in~\cite{sopuerta-yunes-DCS-EMRI,Canizares:2012is} using the semirelativistic approximation, while those for comparable mass binaries were derived in~\cite{Yagi:2012vf} within the PN approximation. Scalar radiation during the merger phase of BH binaries have recently been studied in~\cite{Okounkova:2017yby}.

 GWs from BH binaries whose amplitude being too small to be detected individually form stochastic gravitational wave background (GWB) signals (see~\cite{Allen:1997ad,Romano:2016dpx} for reviews on this topic,~\cite{TheLIGOScientific:2016dpb} for the bound from LIGO's O1 run and \cite{yang2017testing} for measurements using pulsar scintillations). Advanced LIGO may detect such a signal in the near future depending on the averaged mass of BH binaries and their merger rate~\cite{Zhu:2011bd,TheLIGOScientific:2016wyq,Callister:2016ewt}. One can probe parity violation with GWBs by looking for the Stokes \emph{V-mode} parameter in circular polarization. A formalism for detecting circular polarization with GWBs has been developed by Seto~\cite{Seto:2006hf,Seto:2006dz} for space-borne interferometers and by Seto and Taruya~\cite{Seto:2007tn,Seto:2008sr} for ground-based interferometers. Crowder {\it et al.}~\cite{Crowder:2012ik} carried out a Bayesian parameter estimation study based on~\cite{Mandic:2012pj} and derived projected bounds on the V-mode polarization from GWBs with aLIGO. Although these studies had primordial GWBs in mind, their formalisms and calculations are generic that they are also applicable to astrophysical GWBs. 

There are other related works on this topic as follows.
The sensitivities for the V-mode circular polarization with pulsar timing arrays and the cosmic microwave background were calculated in~\cite{Kato:2015bye} and~\cite{Smith:2016jqs} respectively. Interestingly, the former have zero sensitivity for an isotropic background.
Recently, SPIDER placed constraints on the amount of V-mode circular polarization on the cosmic microwave background~\cite{Nagy:2017csq}.
One can also test GR with stochastic GWB signals using ground-based GW detectors by looking for non-tensorial GW polarizations~\cite{Nishizawa:2009bf,Isi:2015cva,Isi:2017equ,Callister:2017ocg} or probing the graviton mass~\cite{Nishizawa:2013eqa}. Maselli {\it et al.}~\cite{Maselli:2016ekw} considered probing non-GR corrections to the GW amplitude with astrophysical GWBs.

\subsection{Goal \& Methodology}

In this paper, we derive bounds on gravitational parity violation with existing LVC events and triggers, and also study how such bounds can be improved in future. We consider both GWs from individual stellar-mass BH binaries and stochastic GWBs created by  such binaries. The former allow us to probe parity violation at each source redshift, while the latter allow us to probe the integrated history of the cosmological evolution of parity violation. Thus, one can probe two different aspects of gravitational parity violation with these two different analyses. We choose CS gravity as an example of a parity-violating theory of gravity and derive projected GW bounds on the local evolution of the scalar field (such as the first time derivative of the local scalar field at present). We work within the weak CS approximation to ensure that the theory is well-posed~\cite{Delsate:2014hba}.

Regarding bounds from individual sources, we carry out a Fisher analysis where we assume that the observed waveforms are consistent with GR. Such an analysis is much simpler to perform than carrying out a full Bayesian parameter estimation study with available data (which we leave for future work). The analysis is similar to what has been done in~\cite{Yunes:2016jcc} except that it now includes corrections to the waveform amplitude. As real data is not needed in such an analysis, the results only depend on what parameters we assume for the injection. Since gravitational birefringence affects the amplitude of GWs where the largest uncertainties come from the sky location and the binary orientation, we perform Monte Carlo simulations within which source positions and orientations are randomly distributed to derive probability distributions of the upper bounds on gravitational parity violation. 

Regarding bounds from GWBs, we first derive the GW energy density spectrum for the V-mode induced by gravitational parity violation. We then apply the technique developed in~\cite{Seto:2007tn,Seto:2008sr} to separate such a mode from the intensity mode, which is insensitive to parity violation. Following~\cite{TheLIGOScientific:2016wyq,Callister:2016ewt}, we assume that the history for binary BH coalescences follows  the formation rate of stars with metallicity smaller than half of the solar metallicity. We adopt the recent estimate of binary BH merger rate $\sim 55$Gpc$^{-3}$yr$^{-1}$, which is based on the LVC observations~\cite{TheLIGOScientific:2016pea}.
 By setting the threshold signal-to-noise ratio (SNR) to be 5 and assuming non-detection of such a V-mode spectrum in future observations, we derive projected constraints on the amount of gravitational parity violation using a network of second-generation detectors or a network of more advanced detectors, such as Voyager and Cosmic Explorer (CE). 
We checked our analysis against a Bayesian parameter estimation study on a simpler model of parity violation in~\cite{Crowder:2012ik} and found that the former qualitatively agrees with the latter and the difference is only $\sim 30\%$. Thus our analysis should give a correct order of magnitude estimate for bounds on CS gravity.

\subsection{Executive Summary}

\begin{figure}[htb]
\includegraphics[width=8.5cm]{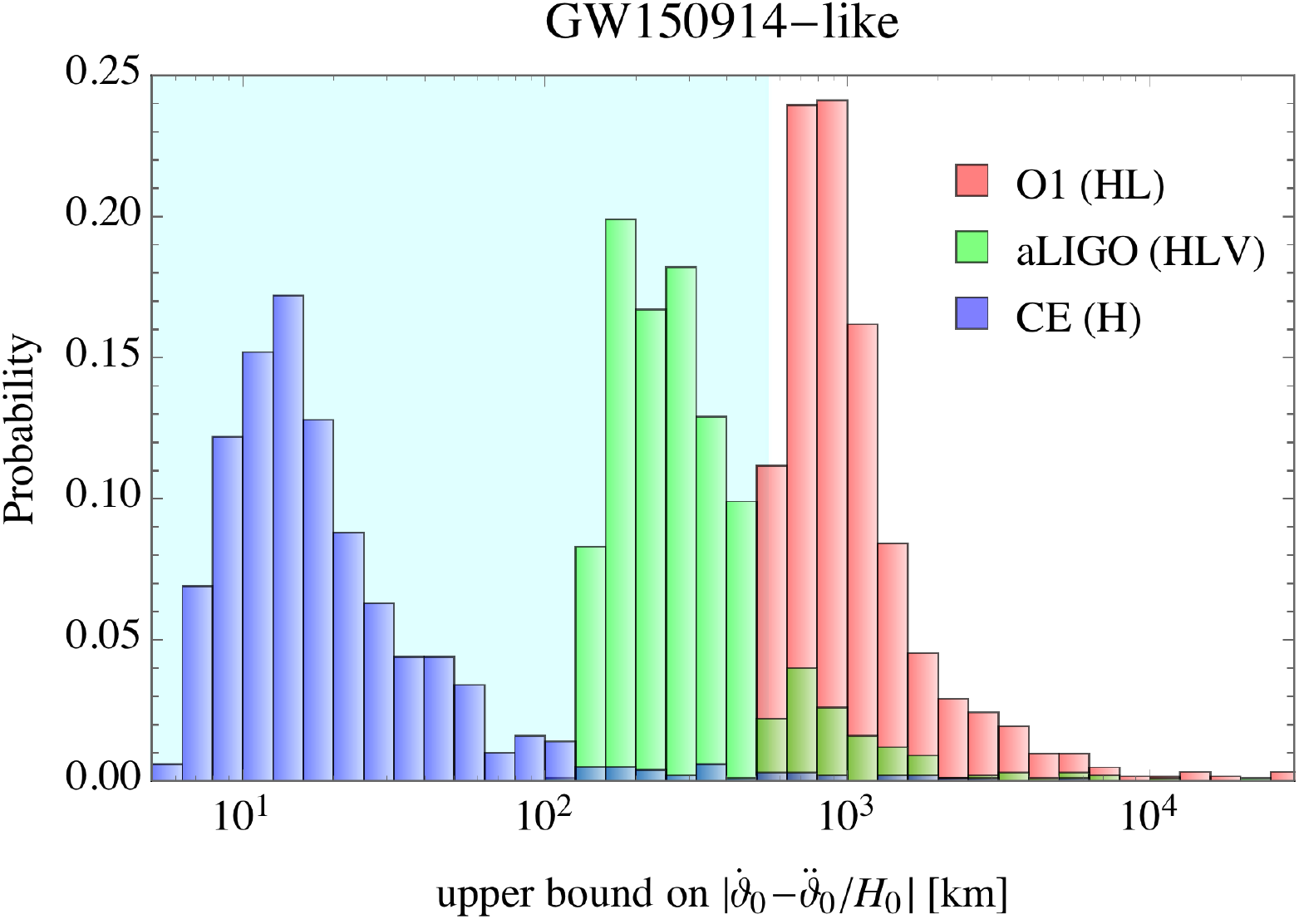}
\caption{Probability distribution of the 1-$\sigma$ upper bounds on the local scalar field evolution in CS gravity for three different types of GW interferometers assuming GW150914-like binaries and randomly distributing the sky location and binary orientation.  
For two aLIGO detectors with the sensitivity at their O1 run (red), we fix the SNR as 23.7 (by adjusting the luminosity distance for a given sky location and binary orientation) and restrict the distance in the range $240 \mrm{Mpc} < D_L < 570 \mrm{Mpc}$ corresponding to the GW150914 measurement.
For a network of three second-generation GW interferometers with their design sensitivity (green) and a third-generation interferometer of CE (blue), we fix the luminosity distance of the source to $D_L = 420$Mpc. 
The weak CS approximation is valid only in the blue shaded region. Observe that the current bounds from the O1 run are too weak to satisfy the approximation, while one should be able to place meaningful bounds in future using a network of second-generation detectors with their design sensitivity or CE.
Bounds from solar system experiments and binary pulsar observations have been derived only on $\dot \vartheta_0$, and these GW observations are likely to place complementary bounds on additional CS parameters.
}
\label{fig:MC-bound}
\end{figure}

Let us now summarize the results for bounds on parity violation using GWs from individual sources. The red histogram in Fig.~\ref{fig:MC-bound} presents the distribution for the 1-$\sigma$ upper bound on the local scalar field evolution $|\dot \vartheta_0 - \ddot \vartheta_0/H_0|$ in CS gravity using GW150914. Here $\vartheta$ represents the scalar field with the subscript ``0'' referring to the local value and a dot represents a derivative with respect to time.
The bounds are valid only when they satisfy the weak CS approximation (blue shaded region). 
Observe that the current bounds from GW150914 are too weak to satisfy the approximation. On the other hand, if GWs from GW150914-like sources are detected by future detectors such as aLIGO with their design sensitivity or CE, the bounds are likely to satisfy the weak CS approximation. Thus, such future detectors allow us to place meaningful bounds on the local scalar field evolution in the theory.

We note that the bounds on $|\dot \vartheta_0-\ddot \vartheta_0/H_0|$ cannot  be directly compared to existing bounds from solar system experiments or binary pulsar observations, as the latter two were derived within the assumption of $\ddot \vartheta = 0$. However, even if one relaxes the assumption of $\ddot \vartheta = 0$, it is likely that such experiments or observations probe different combinations of $\dot \vartheta_0$ and $\ddot \vartheta_0$. This is because the amplitude birefringence effect in GW observations probe the difference in the scalar field derivative between now and at the source redshift (and thus such observations are sensitive to not only $\dot \vartheta_0$ but also $\ddot \vartheta_0$), which is generically different from what solar system experiments or binary pulsar observations are probing. Thus, GW observations should give us complementary bounds on the scalar field evolution in parity-violating gravity compared to other existing experiments or observations.

\begin{figure}[htb]
\includegraphics[width=8.5cm]{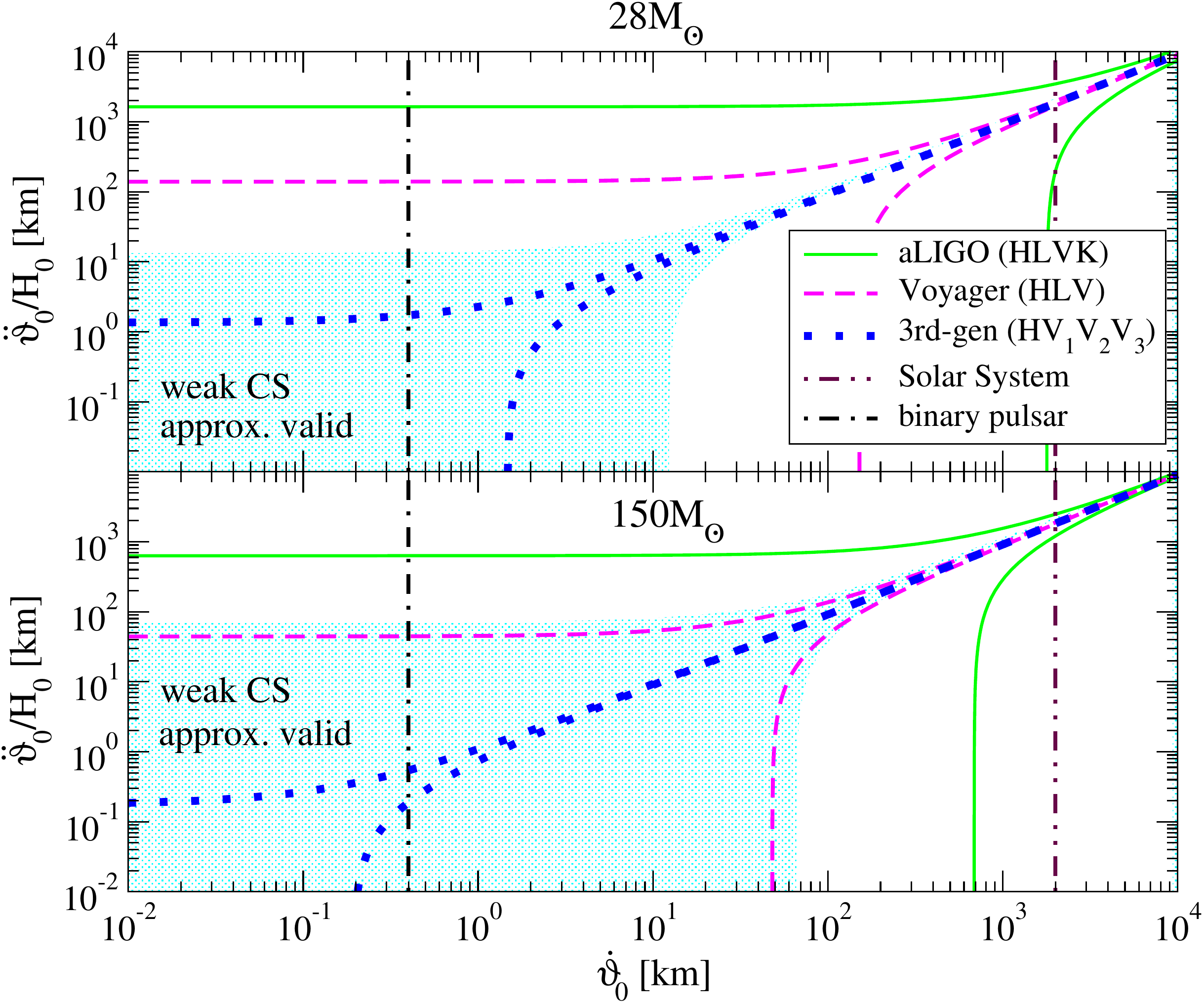}
\caption{Upper bounds on the local evolution of the scalar field $\dot \vartheta_0$ and $\ddot \vartheta_0/H_0$ in CS gravity from stochastic GWBs of stellar-mass BH binaries with the average chirp mass of $28M_\odot$ (top) and $150M_\odot$ (bottom). We assume using a network of four second-generation ground-based GW interferometers (Hanford, Livingston, Virgo, KAGRA) (green solid) and three advanced interferometers (Voyager) (magenta dashed) and two third-generation detectors (CE and ET) (blue dotted). The bounds are derived within the weak CS approximation and are valid only in the blue shaded region. The vertical lines correspond to the bound from the solar system (LAGEOS) experiment~\cite{Smith:2007jm} and binary pulsar observations~\cite{Yunes:2008ua,AliHaimoud:2011bk}, which have only been derived on $\dot \vartheta_0$. Observe that bounds with third-generation detectors can be complementary to the existing bounds and yet satisfy the weak CS approximation.
}
\label{fig:thetadot-thetaddot-pos}
\end{figure}

Let us next explain bounds on parity violation from future GWB observations. Figure~\ref{fig:thetadot-thetaddot-pos} presents the upper bounds on $\dot \vartheta_0$ and $\ddot \vartheta_0/H_0$ in CS gravity with such observations. We assume that the scalar field evolution is given by $\dot \vartheta = \dot \vartheta_0 + \ddot \vartheta_0 t$ with $t=0$ representing the current time. We choose the average chirp mass of the stellar-mass BH binaries constructing stochastic GWBs to be 28$M_\odot$ (top) and $150M_\odot$ (bottom). The former corresponds to that of GW150914 while a larger chirp mass like the latter may be favored if binary BHs are formed in low metallicity environment.
We consider using a network of four second-generation detectors (green solid), three Voyager-type detectors (magenta dashed) or two third-generation detectors corresponding to CE and Einstein Telescope (ET) (blue dotted). 

Several important points can be understood from Fig.~\ref{fig:thetadot-thetaddot-pos}. First, observe that in order to satisfy the weak CS approximation shown by the blue shaded region, one needs third-generation detectors if the average chirp mass is relatively small, while Voyager-type detectors may be able to place meaningful bounds if the average chirp mass is relatively large. Second, observe that these GW bounds are complementary to existing bounds from solar system experiments and binary pulsar observations. Third, notice that the bounds with third-generation detectors are stronger than those with second-generation detectors by $\sim 3$ orders of magnitude, which is much larger than the improvement on the bounds from individual sources (see Fig.~\ref{fig:MC-bound}). This is because the SNR for GWs from individual sources scales linearly with the detector sensitivity while that for stochastic GWBs scales quadratically. Thus, GWB bounds are more sensitive to detector upgrades than bounds from individual sources.

The bounds in  Fig.~\ref{fig:thetadot-thetaddot-pos}  were derived assuming that all source parameters (such as the average BH mass and spin) and the binary BH merger rate  are known \textit{a priori}, and thus correspond to  an ideal situation. In a more realistic situation where such information is no longer available, approximate bounds can be derived if the GW energy density spectrum of the regular intensity mode (or the Stokes parameter's I-mode) is measured. In such a situation, the only a priori information required is the peak redshift of the binary BH merger-rate history. If such a merger rate follows the star formation rate history, the peak redshift is $\sim 1.5$. With these pieces of  information at hand, we have derived approximate bounds on $\dot \vartheta_0$ and $\ddot \vartheta_0$, which are accurate for the smaller average chirp mass case and may deviate from the true value by a factor of $\sim 3$ when the average chirp mass is relatively large. Such a finding shows that the approximate bounds are still valid as  order of magnitude estimates.

\subsection{Organization}
This paper is organized as follows. In Sec.~\ref{sec:birefringence}, we introduce GW amplitude birefringence in gravitational parity violation. We consider CS gravity as an example. In Sec.~\ref{sec:individual-source}, we derive bounds on such parity violation with GWs from individual sources via a Fisher analysis. In Sec.~\ref{sec:GWB}, we study bounds with stochastic GWBs from stellar-mass BH binaries. Finally in Sec.~\ref{sec:conclusion}, we discuss possible avenues for future work.

\section{Amplitude Birefringence in Gravitational Parity Violation}  
\label{sec:birefringence}

In this section, we review the amplitude birefringence effect in CS gravity. After describing the basics of CS gravity in Sec.~\ref{sec:CS}, we explain how amplitude birefringence occurs in this theory in Sec.~\ref{sec:birefringence2}. 

\subsection{ABC of Chern-Simons Gravity}
\label{sec:CS}
In this section, we review amplitude birefringence of GWs in a parity-violating theory of gravity.
As an example, we work on CS gravity~\cite{jackiw,CSreview}. The action is given by
\ba
S &\equiv & \int d^4x \sqrt{-g} \left\{ \frac{R}{16\pi} + \frac{\vartheta}{64\pi}  R\, {}^* \! R \right.  \nn \\
& & \left.  - \frac{\beta}{2} \left[ \nabla_\mu \vartheta \nabla^{\mu} \vartheta + 2 V(\vartheta) \right] 
+ \mathcal{L}_{\MAT} \right\}\,,
\label{action}
\ea
where $g$ and $R$ are the metric determinant and Ricci scalar respectively while $\vartheta$ is the scalar field (with the unit of length squared) 
with a potential $V$ 
and $\mathcal{L}_{\MAT}$ is the matter Lagrangian density. $\beta$ is a dimensionless parameter with $\beta =1$ representing the theory with the canonical scalar field while $\beta=0$ representing the non-dynamical formulation. $R\, {}^* \! R$ is the Pontryagin density defined via
\be
R\, {}^* \! R \equiv \frac{1}{2} R_{\alpha \beta \gamma \delta} \varepsilon^{\alpha \beta \mu \nu} R^{\gamma\delta}{}_{\mu\nu}\,,
\ee
with $\varepsilon^{\alpha \beta \mu \nu}$ representing the Levi-Civita tensor.

The modified field equations are given by
\be
\label{eq:mod-Ein}
G_{\mu\nu} + 16\pi C_{\mu\nu} =8\pi \left( T_{\mu\nu}^{(\vartheta )} + T_{\mu\nu}^\mrm{mat} \right)\,,
\ee
where the $C$-tensor is defined by
\be
\label{eq:C-tensor}
C^{\mu\nu}  \equiv  (\nabla_\sigma \vartheta) \epsilon^{\sigma\delta\alpha(\mu} \nabla_\alpha R^{\nu)}{}_\delta + (\nabla_\sigma \nabla_\delta \vartheta) {}^* R^{\delta (\mu\nu) \sigma}\,,
\ee
with bracket on indices denoting the symmetrization operation and $ {}^* R_{\delta \mu\nu \sigma}$ representing the dual of the Riemann tensor. $T_{\mu\nu}^{(\vartheta )}$ is the stress-energy tensor for the scalar field defined by
\be
T_{\mu\nu}^{(\vartheta )} \equiv \beta (\nabla_\mu \vartheta) (\nabla_\nu \vartheta) - \frac{\beta}{2} g_{\mu\nu} [\nabla_\delta \vartheta \nabla^{\delta} \vartheta + 2 V(\vartheta)]\,,
\ee
while $T_{\mu\nu}^\mrm{mat}$ is the stress-energy tensor for the matter field.
The scalar field equation is given by
\be
\beta\, \square \vartheta = - \frac{1}{64 \pi} R\, {}^* \! R + \beta \frac{dV}{d\vartheta}\,.
\ee
A Friedmann-Robertson-Walker (FRW) spacetime is also a solution to CS gravity as such a spacetime leads to a vanishing Pontryagin density~\cite{Guarrera:2007tu}. In such a case, the scalar field also obeys the background metric symmetry, namely homogeneous and isotropic.

As already mentioned in the introduction, we will treat this theory as an effective field theory. This is because the C-tensor in Eq.~\eqref{eq:C-tensor} contains third derivatives and the theory is not well posed in terms of the initial value problem~\cite{Delsate:2014hba}. We will work in the weak CS approximation which is valid as long as CS corrections are always smaller than GR terms in the equation of motion. 
%

\subsection{Amplitude Birefringence}
\label{sec:birefringence2}

We next consider a (comoving) linear metric perturbation 
\be
h_{ij} = a(\eta)\, \bar h_{ij}(\eta,\chi^i)
\ee
under a FRW spacetime given by
\be
ds^2 = a^2(\eta) \left[ -d\eta^2 +(\delta_{ij} + \bar h_{ij}) d\chi^i d\chi^j \right]\,,
\ee 
where $\eta$ is the conformal time, $\chi^i$ is the comoving spatial coordinates and $a$ is the scale factor. We assume that the scalar field $\phi$ is a function of $\eta$ only to ensure that the background symmetry is preserved.
As in GR, such a symmetry further ensures that one can decompose perturbations into scalar, vector and tensor~\cite{Dyda:2012rj}  components that decouple. Thus, we only consider tensor perturbations. 
The linearized modified Einstein equations in the transverse and traceless gauge are given by~\cite{Yunes:2010yf} (see also~\cite{Alexander:2007kv,Alexander:2004wk})
\be
\label{eq:lin-Ein-eq}
\square_g \bar h^j_i = -\frac{1}{a^2} \varepsilon^{pjk}\left[ \left( \vartheta''-2\mathcal{H}\vartheta' \right) \partial_p \bar h_{ki}'+\vartheta' \partial_p \square_g \bar h_{ki} \right]\,,
\ee
where a prime denotes a conformal time derivative, $\mathcal{H} \equiv a'/a$ is the conformal Hubble parameter and
\be
\square_g \equiv \partial_\eta^2 - \partial_i \partial^i + 2 \mathcal{H} \partial_\eta\,.
\ee
Notice that Eq.~\eqref{eq:lin-Ein-eq} does not depend on $\beta$, and hence the analysis here is valid for both dynamical and non-dynamical formulation.

We next decompose $\bar h_{ij}$ in terms of different polarization states:
\be
\label{eq:h-decomp}
\bar h_{ij} = \sum_P \bar h_P e_{ij}^P\,.
\ee
Here, $P$ denotes polarization states and $e_{ij}^P$ is the polarization basis. A common choice is the $+$ and $\times$ mode polarizations, but circular polarizations are more useful when probing parity violation. Thus, we choose $P=(R,L)$ which corresponds to the right-handed and left-handed modes respectively. The circular polarization bases are connected to the $+$ and $\times$ mode polarization tensors $e_{ij}^{+}$ and $e_{ij}^{\times}$ as (see e.g.~\cite{Seto:2008sr})
\be
e_{ij}^{R} = \frac{e_{ij}^+ + i e_{ij}^{\times}}{\sqrt{2}}\,, \quad e_{ij}^{L} = \frac{e_{ij}^+ - i e_{ij}^{\times}}{\sqrt{2}}\,,
\ee
which obeys
\be
\label{eq:pol-tensor}
\varepsilon^{ijk}n_i e_{kl}^{R,L} = i \lambda_{R,L} e^j{}_l{ }^{R,L}\,,
\ee
with $\lambda_R=+1$ and $\lambda_L=-1$.
Similarly, one can write $\bar h_{R,L}$ in terms of $\bar h_{+,\times}$ as
\be
\bar h_{R} = \frac{\bar h_+ - i \bar h_{\times}}{\sqrt{2}}\,, \quad \bar h_{L} = \frac{\bar h_+ + i \bar h_{\times}}{\sqrt{2}}\,.
\ee
We further decompose $\bar h_{R,L}$ as 
\be
\bar h_{R,L} = \mathcal{A}_{R,L} e^{-i\left[ \phi(\eta)-\kappa n_k \chi^k \right]}\,.
\ee
Here $\mathcal{A}_{R,L}$ is the amplitude, $\phi$ is the phase, $\kappa$ is the conformal wave number and $n^k$ is the unit vector representing the direction of the wave propagation.

Let us now look at the dispersion relation.
Substituting Eq.~\eqref{eq:h-decomp} to Eq.~\eqref{eq:lin-Ein-eq}, one finds the following relation~\cite{Yunes:2010yf}:
\be
\label{eq:disp-rel}
i \phi'' + (\phi')^2 - \kappa^2 = -2i \frac{\mathcal S_{R,L}'}{\mathcal S_{R,L}} \phi'\,,
\ee
where
\be
\mathcal S_{R,L} \equiv a \sqrt{1-\lambda_{R,L} \frac{\kappa \vartheta'}{a^2}}
\ee
corresponds to the effective scale factor in CS gravity. 
Let us now impose the requirement that
\be
(\phi')^2 \gg \phi''\,, 
\ee
together with
\be
\label{eq:approx}
\kappa \gg \frac{\mathcal{S}'_{R,L}}{\mathcal{S}_{R,L}}\,.
\ee
Eq.~\eqref{eq:approx} can be satisfied by using $\kappa \gg \mathcal{H}$ (GW wavelengths much shorter than the Hubble scale) and imposing the weak CS approximation:
\be
\label{eq:weak-CS}
\kappa |\vartheta'| \ll a^2\,, \quad \kappa |\vartheta'' | \ll 2 a^2 \mathcal{H}\,,
\ee
which can be rewritten as
\be
\label{eq:weak-CS2}
 |\dot \vartheta| \ll \frac{1}{2 \pi (1+z) f}  \,, \quad
 \bigg| \dot \vartheta + \frac{\ddot \vartheta}{H} \bigg| \ll \frac{1}{\pi (1+z) f} \,.
\ee

One can then solve the dispersion relation in Eq.~\eqref{eq:disp-rel} to yield~\cite{Yunes:2010yf}
\be
\phi_{R,L}(\eta)=\pm \kappa (\eta-\eta_s)+i \ln \left[ \frac{\mathcal S_{R,L}(\eta_s)}{\mathcal S_{R,L}(\eta)} \right]\,,
\ee
where $\eta_s$ is the conformal time at which GWs are emitted.
Imposing further Eq.~\eqref{eq:weak-CS} and setting $\eta = 1$ (the present conformal time), one finds 
\be
\label{eq:phase-eta0}
\phi_{R,L}(1) = \pm \kappa (1-\eta_s)+i \lambda_{R,L} \pi f \dot \Theta\,.
\ee
Here $f \equiv \kappa/(2\pi a_0)$ is the (observed) GW frequency with the subscript 0 representing a quantity to be evaluated at $\eta=1$ while
\be
\label{eq:Thetadot}
\dot \Theta \equiv  \dot \vartheta_0 -(1+z) \dot \vartheta_s
\ee
with  the subscript $s$ representing a quantity to be evaluated at $\eta=\eta_s$ and $z$ representing the source redshift defined by $z \equiv a_0/a_s-1$. A dot refers to a derivative with the physical time $t$ given by $t = \int a d\eta$.

Notice that the CS correction in the phase (the second term in Eq.~\eqref{eq:phase-eta0}) is purely imaginary, which means that it enters as an amplitude modulation to the gravitational waveform. Depending on the sign of $\dot \Theta$, one circular polarization is amplified while the other polarization is suppressed during the wave propagation. Notice also that such amplitude birefringence is absent in GR ($\vartheta \to 0$).
Having the above result at hand, one can decompose $\bar h_{R,L}$ into the GR and CS contribution as
\be
\label{eq:hv}
\bar h_{R,L} = \bar h_{R,L}^\GR \left( 1+\lambda_{R,L} v \right)
\ee
with
\be
\label{eq:v}
v \equiv \pi f \dot \Theta
\ee
representing the relative gravitational parity violation in the waveform.

\section{Gravitational Waves from Individual Sources}  
\label{sec:individual-source}

In this section, we describe bounds on parity violation with GWs from individual BH binaries. We first explain how the gravitational waveform is modified from GR. We next describe the detector sensitivity and how one can carry out a Fisher analysis to derive the bounds. We end this section by presenting and interpreting the results. 
 
\subsection{Gravitational Waveform}

As we will see in the next subsection, GW data analysis is done in the Fourier domain. The Fourier waveform $\tilde h (f)$ consists of the $+$ and $\times$ mode waveform~\cite{Veitch:2009hd}:
\ba
\label{eq:h}
\tilde h (f) & =& \left[F_+(\theta_d, \phi_d, \psi_d) \, \tilde h_+ (f)  + F_\times (\theta_d, \phi_d, \psi_d) \, \tilde h_\times (f) \right] \nn \\
& & \times e^{-2\pi i f \Delta t}\,.
\ea
Here $F_+$ and $F_\times$ are the beam pattern functions (see e.g.~\cite{Sathyaprakash:2009xs}) that depend on the source location (polar angle $\theta_d$ and azimuthal angle $\phi_d$) and polarization angle $\psi_d$ in the detector frame while $\Delta t(\theta_d, \phi_d)$ is the arrival time difference between the detector and the geocenter. 

We first review the waveform in GR. $\tilde h_+$ and $\tilde h_\times$ are given by 
\be
\label{eq:h-GR}
\tilde h_+^\GR = (1+\mu^2) A  e^{i\Psi}\,, \quad \tilde h_\times^\GR = 2 \, \mu \, A e^{i(\Psi + \pi/2)}
\ee
with  $A$ and $\Psi$ representing the amplitude and phase of the GR Fourier waveform and $\mu=\cos \iota$ where $\iota$ is the inclination angle. In this paper, we use the inspiral-merger-ringdown phenomenological B (IMRPhenomB) waveform constructed by fitting numerical relativity waveforms of binary BH coalescences\footnote{A more up-to-date phenomenological waveform (IMRPhenomD) is also available, though the systematics due to the difference between the two IMRPhenom waveforms are much smaller than statistical errors on non-GR parameters entering in the GW propagation~\cite{Yunes:2016jcc}.}. The amplitude and phase can be found in Eq.~(1) of~\cite{Ajith:2009bn} with $\mathcal{C} = \sqrt{5/96} \mathcal{M}_z/(\pi^{2/3} D_L)$ (see e.g.~\cite{Taylor:2012db}), where $\mathcal{M}_z$ is the redshifted chirp mass that we shall define in the next subsection while $D_L$ is the luminosity distance. 
 
We now derive corrections to the waveform in CS gravity. $\tilde h_+$ and $\tilde h_\times$ are given in terms of the GR waveform as~\cite{Yunes:2010yf}
\be
\label{eq:h-CS}
\tilde h_+ = \tilde h_+^\GR - i v \tilde h_\times^\GR\,, \quad \tilde h_\times = \tilde h_\times^\GR + i v \tilde h_+^\GR
\ee
with $v$ given in Eq.~\eqref{eq:v}. 
Substituting Eqs.~\eqref{eq:h-GR} and~\eqref{eq:h-CS} into Eq.~\eqref{eq:h}, one finds
\be
\tilde h (f) = A\, \delta A\, e^{i (\Psi + \delta \Psi)}\,,
\ee
where\footnote{This corrects typos in~\cite{Yunes:2013dva}.} 
\ba
\allowdisplaybreaks
\label{eq:amp-corr}
\delta A &=&\sqrt{
   \left(1+\mu ^2 + 2 \mu  v\right)^2 F_+^2+ \left[2 \mu + (1+\mu ^2) v\right]^2 F_\times^2} \nn \\
 &=& \sqrt{(1+\mu^2)^2 F_+^2 + 4 \mu^2 F_\times^2} \nn \\
& &\times \left[1+\frac{2 \mu (1+\mu^2) (F_+^2 + F_\times^2)}{(1+\mu^2)^2 F_+^2 + 4 \mu^2 F_\times^2}v + \mathcal{O}(v^2) \right]\,, \\
\label{eq:phase-corr}
\delta \Psi &=& \tan ^{-1}\left\{\frac{  \left[2 \mu + (1+\mu ^2) v\right] F_\times}{ \left(1+\mu ^2+2 \mu  v\right) F_+}\right\} \nn \\
&= & \tan^{-1}\left[ \frac{2\mu F_+}{(1+\mu^2) F_\times} \right] + \frac{(1-\mu^2)^2 F_+ F_\times}{(1+\mu^2)^2F_+^2 + 4 \mu^2 F_\times^2} v \nn \\
&& + \mathcal{O}(v^2)\,.
\ea
The GR contribution to $\delta A$ and $\delta \Psi$ agree with those in~\cite{cutler1998,Berti:2004bd}. Such a waveform correction can be mapped to parameterized post-Einsteinian waveform~\cite{PPE,Yunes:2013dva} that captures non-GR modifications in the waveform in a generic way.

Let us now count the PN order of the above parity-violation corrections relative to GR. The relative correction from GR is said to be of $n$~PN order if it is proportional to $f^{2n/3}$. Thus, the amplitude correction in Eq.~\eqref{eq:amp-corr} enters at 1.5PN order. On the other hand, since $\Psi \propto f^{-5/3}$ at leading order, the phase correction in Eq.~\eqref{eq:phase-corr} enters at 4PN order. The latter enters at the same PN order as the time of coalescence and these two parameters are strongly degenerate. Thus, the CS correction is mainly constrained from the amplitude correction. Since the IMRPhenomB waveform includes up to 1.5PN terms in the inspiral part of the amplitude, our analysis takes into account correlations between the GR and CS correction terms in the amplitude entering at the same PN order.  

\subsection{Data Analysis Formalism}

One can estimate parameter uncertainties in GW observations via a Fisher analysis~\cite{Finn:1992wt,cutlerflanagan}, which is valid for sufficiently large SNR events. For a Gaussian and stationary noise, the posterior distribution of a set of parameters $\theta^a$ with a given measurement data $s$ is given by
\be
p({\theta}^{a}|s) \propto p^{(0)}({\theta}^{a}) \exp\left[ - \frac{1}{2} \Gamma_{ab} \left( \theta^a - \hat \theta^a \right) \left( \theta^b - \hat \theta^b \right) \right]\,,
\ee
where $p^{(0)}({\theta}^{a})$ is the prior distribution of parameters while the exponential part is the likelihood distribution and $\hat \theta^a$ is the maximum likelihood value of each parameter. The Fisher matrix is defined by 
\be
\label{eq:Fisher}
\Gamma_{ab} \equiv (\partial_a h | \partial_b h)\,,
\ee
where $\partial_a h \equiv \partial h / \partial \theta^a$ and the inner product is defined by
\be
\label{eq:inner}
(A|B) \equiv 4 \mrm{Re} \int_0^\infty df \frac{\tilde A^* (f) \tilde B(f)}{S_n(f)}\,.
\ee
Here the superscript * denotes the complex conjugate and $S_n(f)$ is the noise spectral density.
In practice, the integral in Eq.~\eqref{eq:inner} is calculated from a minimum frequency $f_\mrm{min}$ to a maximum one $f_\mrm{max}$, which shall be discussed in more detail later. One can define the SNR using the above inner product as
\be
\rho^2 = (h|h)\,.
\ee

Following~\cite{cutlerflanagan,Poisson:1995ef,Berti:2004bd}, we introduce the prior in a rather crude way by assuming that it has a Gaussian distribution centered around $\bar \theta^a$ (whose choice is irrelevant for estimating statistical errors) with variance $\sigma_{\vartheta^a}^2$:
\be
\label{eq:prior}
p^{(0)} \propto \exp\left[ - \frac{1}{2} \sum_a \left( \frac{\theta^a - \bar{\theta}^a}{\sigma_{\theta^a}}\right)^2 \right]\,.
\ee
Using the fact that the product of two Gaussian is also a Gaussian, the standard deviation of $\theta^a$ is then given by
\be
\Delta \theta^a = \sqrt{\left(\tilde \Gamma^{-1} \right)_{aa}}\,, 
\qquad 
\tilde \Gamma_{ab} \equiv \Gamma_{ab} + \frac{1}{\sigma_{\theta^a}^2} \delta_{ab}\,. 
\ee

In this paper, we choose the parameters as 
\be
\label{eq:par-ground}
\theta^a = (\ln \mathcal{M}_z, \ln \bar \eta, \chi, t_c, \phi_c, \ln D_L, \alpha,\delta,\psi,\iota, \dot \Theta)\,.
\ee
Here, $\bar \eta \equiv m_1 m_2/(m_1+m_2)^2$ is the symmetric mass ratio with $m_A$ representing the mass of the $A$th body, $\mathcal{M}_z \equiv (1+z) (m_1+m_2) \bar \eta^{3/5}$ is the redshifted chirp mass with the redshift $z$, $\chi \equiv (m_1 \chi_1+m_2 \chi_2)/(m_1+m_2)$ is the effective spin parameter~\cite{Ajith:2009bn} with $\chi_A$ representing the dimensionless spin of the $A$th body, $t_c$ and $\phi_c$ are the coalescence time and phase, $D_L$ is the luminosity distance, $\alpha$, $\delta$ and $\psi$ are the right ascension, declination and polarization angle in the Earth fixed frame while $\iota$ is the inclination angle. 
We choose the injection masses and the luminosity distance (or SNR) as those corresponding to GW150914 summarized in Table~\ref{table:events}. For other parameters, we use $\chi=0$, $t_c = \phi_c = 0$ and $\dot \Theta = 0$. We uniformly distribute the sky location $(\alpha, \delta)$ and the binary orientation $(\psi,\iota)$. We impose the prior on spin, coalescence phase, sky location and binary orientation following Eq.~\eqref{eq:prior} with $\sigma_{\chi} = 1$, $\sigma_{\phi_c} = \pi$, $\sigma_{\alpha}=\pi$, $\sigma_{\delta}=\pi/2$, $\sigma_{\psi}=\pi$, $\sigma_{\iota}=\pi/2$ \footnote{Such priors slightly help to break degeneracies among these parameters and $\dot \Theta$. For example, the median and 68\% quantile for the upper bound distribution of $|\dot \vartheta_0 - \ddot \vartheta_0/H_0|$ with O1 (HL) in Table~\ref{table:bounds} increase to 1040km and 1250km respectively if one does not include priors.}. 

{
\newcommand{\minitab}[2][l]{\begin{tabular}{#1}#2\end{tabular}}
\renewcommand{\arraystretch}{1.2}
\begingroup
\begin{table}[htb]
\begin{centering}
\begin{tabular}{ccccc}
\hline
\hline
\noalign{\smallskip}
 SNR & 
 $m_1 [M_\odot]$ & $m_2 [M_\odot]$   & $D_L [\mrm{Mpc}]$ & $z$ \\ \hline
 23.7  & $36.2^{+5.2}_{-3.8}$ & $29.1^{+3.7}_{-4.4}$ & $420^{+150}_{-180}$  & $0.09^{+0.03}_{-0.04}$  \\ 
\noalign{\smallskip}
\hline
\hline
\end{tabular}
\end{centering}
\caption{Parameters for GW150914~\cite{TheLIGOScientific:2016pea}.
}
\label{table:events}
\end{table}
\endgroup
}

\begin{figure}[htb]
\includegraphics[width=8.5cm]{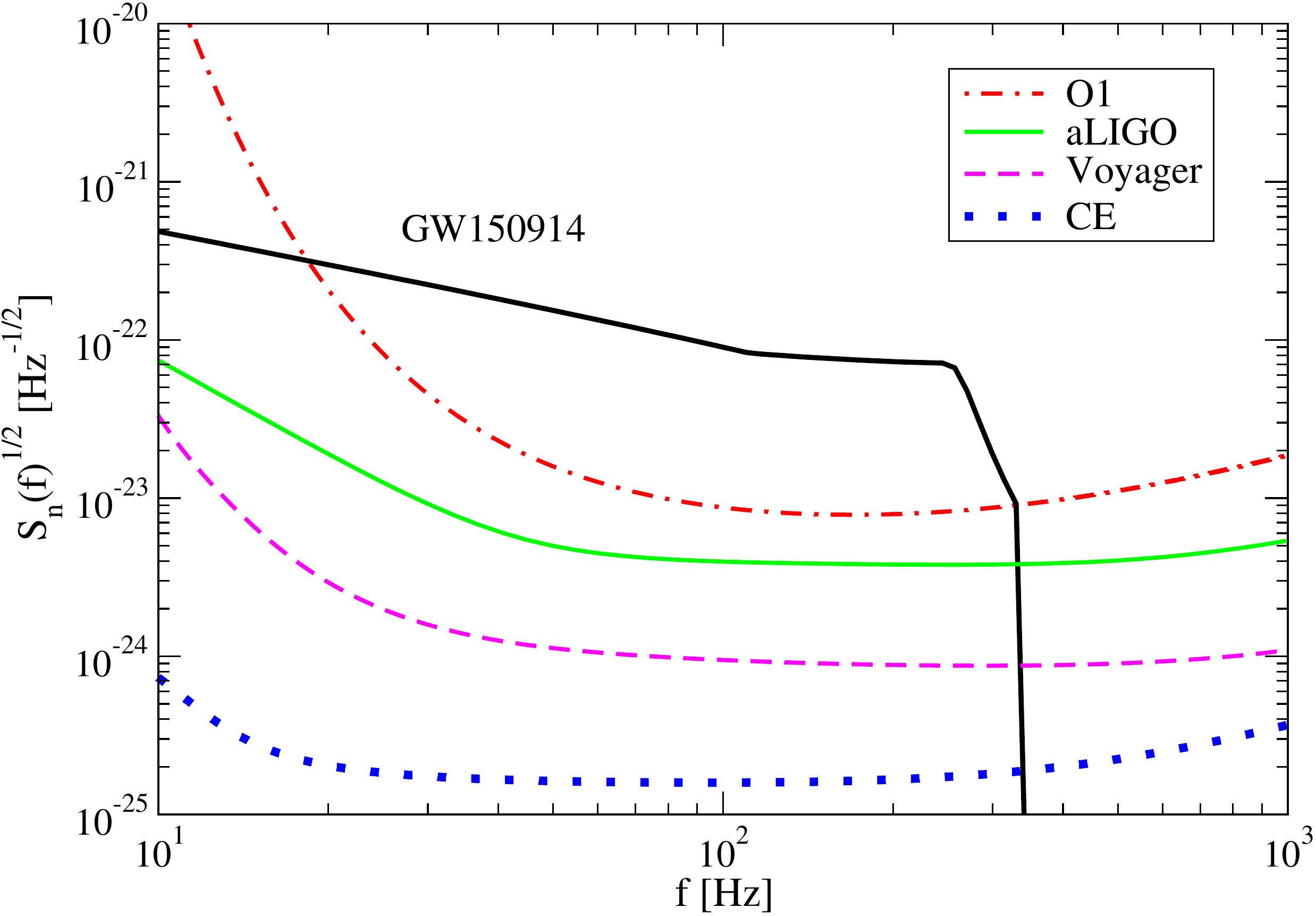}
\caption{Noise spectral density for various interferometers. For reference, we also show the (sky-averaged) GW spectrum $2 \sqrt{f} |\tilde h (f)|$ for GW150914 (thick black solid). The ratio between the GW spectrum and the noise sensitivity curves roughly corresponds to the SNR.}
\label{fig:noise}
\end{figure}

We now explain the detector noise sensitivity. In this paper, we consider both second-generation and third-generation GW interferometers. 
For simplicity we assume that in the former case, detectors at Hanford (H), Livingston (L) and Virgo (V) (and KAGRA (K) in Sec.~\ref{sec:GWB}) all have the same sensitivity. We consider two types of second-generation interferometers: aLIGO O1 run~\cite{noise-data,TheLIGOScientific:2016zmo,Yunes:2016jcc} and aLIGO design sensitivity with zero-detuned, high power configuration~\cite{Ajith:2011ec} (we simply refer to the former as ``O1'' and the latter as ``aLIGO''). We also consider CE~\cite{Evans:2016mbw} as a representative of the third-generation GW interferometers. In Sec.~\ref{sec:GWB}, we also consider Voyager~\cite{adv-det}. The noise spectral density for these GW detectors are presented in Fig.~\ref{fig:noise}.
The minimum and maximum frequency for calculating the Fisher matrix is chosen as $f_\mrm{min}=10$Hz and $f_\mrm{max}=f_\mrm{term}$, where $f_\mrm{term}$ is the terminating frequency of the IMRPhenomB waveform. 

\subsection{Results}

We now explain the results of our Fisher analysis. We begin by considering amplitude birefringence from nearby sources of $z \ll 1$ in general. Then, one can parameterize $\dot \vartheta$ as $\dot \vartheta = \dot \vartheta_0 + \ddot \vartheta_0 t + \mathcal{O}(t^2) = \dot \vartheta_0 - (\ddot \vartheta_0/H_0) z + \mathcal{O}(z^2)$~\cite{Yunes:2010yf}, where we set the current time as $t=0$. 
Substituting this into Eqs.~\eqref{eq:Thetadot} and~\eqref{eq:v}, one finds
\be
v = -\pi \left( \dot \vartheta_0 - \frac{\ddot \vartheta_0}{H_0} \right) f z\,. 
\ee

The red histogram in Fig.~\ref{fig:MC-bound} presents the distribution for the 1-$\sigma$ upper bound on $|\dot \vartheta_0 - \ddot \vartheta_0/H_0|$ for GW150914.
Let us check whether these bounds satisfy the weak CS approximation in Eq.~\eqref{eq:weak-CS} or~\eqref{eq:weak-CS2} that was used to derive the CS corrections to the gravitational waveform. 
In particular, when $z \ll 1$, these conditions become 
\ba
 |\dot \vartheta_0| &\ll& \frac{1}{2 \pi f} = 1.4 \times 10^2 \mrm{km} \left( \frac{350 \mrm{Hz}}{f} \right)  \,, \\
 \bigg| \dot \vartheta_0 + \frac{\ddot \vartheta_0}{H_0} \bigg| &\ll& \frac{1}{\pi f} = 2.7 \times 10^2 \mrm{km} \left( \frac{350 \mrm{Hz}}{f} \right)\,.
\ea
Here $f = 350$Hz corresponds to the termination frequency for GW150914 in the IMRPhenomB waveform~\cite{Ajith:2009bn}. Using the above equations, one finds 
\ba
\bigg| \dot \vartheta_0 - \frac{\ddot \vartheta_0}{H_0} \bigg| &=&  \bigg| 2 \dot \vartheta_0 - \left( \dot \vartheta_0 + \frac{\ddot \vartheta_0}{H_0} \right) \bigg| \nn \\
&\ll& \frac{2}{\pi f} = 5.4 \times 10^2 \mrm{km} \left( \frac{350 \mrm{Hz}}{f} \right)\,.
\ea
The region that satisfies such weak CS approximation is shown by the blue shaded area in Fig.~\ref{fig:MC-bound}. Observe that the bounds with the GW150914 observation do not satisfy the condition.

Let us next study how such bounds improve in future. We consider a network of three second-generation detectors (Hanford, Livingston, Virgo) with each of them having aLIGO's design sensitivity. We also assume that CE is built at the Hanford site. Figure~\ref{fig:MC-bound} compares the bounds with GW150914 using aLIGO's O1 run, aLIGO's design sensitivity and CE (single interferometer). The latter two histograms are obtained by setting the luminosity distance as $D_L = 420$Mpc instead of fixing the SNR. Observe that the weak CS condition is likely to be satisfied with both aLIGO's design sensitivity and CE. Thus one can place meaningful constraints on the local scalar field evolution with such future detectors.
 The mean and 68\% quantile for the upper bound on $|\dot \vartheta_0 - \ddot \vartheta_0/H_0|$ are summarized in Table~\ref{table:bounds}. Observe that the bounds with aLIGO's design sensitivity and CE become stronger than that for the  O1 observation by a factor of $\sim 3$ and $\sim 40$ respectively. Such an enhancement roughly agrees with the relative improvement in $\sqrt{S_n}$ in Fig.~\ref{fig:noise}. 

{
\newcommand{\minitab}[2][l]{\begin{tabular}{#1}#2\end{tabular}}
\renewcommand{\arraystretch}{1.2}
\begingroup
\begin{table}[htb]
\begin{centering}
\begin{tabular}{c|ccc}
\hline
\hline
\noalign{\smallskip}
 &  \multicolumn{1}{c}{O1 (HL)} & aLIGO (HLV) & CE (H) \\  \hline
median & (920)    & 261.5 &18  \\ 
68\% quantile & (1100)   &  338.1 &25 \\ 
\noalign{\smallskip}
\hline
\hline
\end{tabular}
\end{centering}
\caption{Median and 68\% quantile for the upper bound distribution of $|\dot \vartheta_0 - \ddot \vartheta_0/H_0|$ [km] in Fig.~\ref{fig:MC-bound}. Numbers in brackets do not satisfy the weak CS approximation.
}
\label{table:bounds}
\end{table}
\endgroup
}

We now compare the bounds in Fig.~\ref{fig:MC-bound} to existing bounds. A frame-dragging measurement with LAGEOS places the bound $|\dot \vartheta_0 | \lesssim 2000$km for $\ddot \vartheta =0$~\cite{Smith:2007jm}. On the other hand, binary pulsar observations place the bound $|\dot \vartheta_0 | \lesssim 0.4$km, again for $\ddot \vartheta =0$~\cite{Yunes:2008ua,AliHaimoud:2011bk}. Thus, when $\ddot \vartheta =0$, the GW bounds on $\dot \vartheta_0$ are typically stronger (weaker) than the solar system (binary pulsar) one. Although the bounds for $|\dot \vartheta_0 - \ddot \vartheta_0/H_0 |$ with solar system experiments and binary pulsar observations have not been derived yet, GW observations are likely to place complementary bounds in such a combination of parameters. This is because unlike the former, GW observations are sensitive to the \emph{difference} in $\dot \vartheta$ between the source and us (see Eq.~\eqref{eq:Thetadot}), and thus allow us to probe $\ddot \vartheta$ more accurately than other experiments and observations.

\section{Stochastic Gravitational Wave Background}  
\label{sec:GWB}

We next consider bounds on parity violation with stochastic GWB signals originated from stellar-mass BH binaries. We begin by introducing the Stokes parameters, in particular the I- and V-mode polarizations. We next derive the energy density spectrum for each mode. We then review how one can separate these modes with the detector sensitivity of each mode. We end this section by discussing the results.

\subsection{Gravitational-wave Energy Density Spectrum}
\label{sec:GW-spectrum}

We start by expanding a GWB in terms of plane waves with frequency $f$ from a direction $n$. As we explained in Sec.~\ref{sec:birefringence2}, parity violation during propagation only affects the GW amplitude, and hence one can formerly adopt the plane wave decomposition as in GR~\cite{Seto:2007tn,Seto:2008sr}: 
\be
\label{eq:plane-wave}
h_{ij}(t,\vec{x}) = \sum_{P} \int^\infty_{-\infty} df \int d^2 \Omega \, h_P(f,n) e^{-2\pi i f (t-n\cdot \vec{x})} e_{ij}^P(n)\,.
\ee
We recall that $P$ denotes polarization states which we choose $P=(R,L)$ and $e_{ij}^P$ is the polarization basis that obeys Eq.~\eqref{eq:pol-tensor}. We now assume that the stochastic GWB is stationary, Gaussian and isotropic. The quadratic expectation values (ensemble average) for each polarization of the background can be written as~\cite{Seto:2007tn,Seto:2008sr}
\ba
\label{eq:IV}
\begin{pmatrix}
\langle h_R(f,n)\, h_R^*(f',n') \rangle \\
\langle h_L(f,n)\, h_L^*(f',n') \rangle 
\end{pmatrix}
&=&\frac{1}{2} \delta (f-f') \delta^2(n , n') \nn \\
&& \times
\begin{pmatrix}
I(f) + V(f) \\
I(f) - V(f)
\end{pmatrix}\,.
\ea
Here $I$ and $V$ are the Stokes parameters corresponding to the total (squared) amplitude and the asymmetry between the right-handed and left-handed amplitudes respectively. Non-vanishing $V$ signals parity violation as the parity transformation interchanges  two circular polarization modes. $I$ is related to the fractional GW energy density spectrum by~\cite{Flanagan:1993ix,Allen:1997ad}
\be
\label{eq:Omega-I}
\Omega_\GW^{(I)} (f) \equiv \frac{1}{\rho_c} \frac{d\rho_\GW}{d \ln f} = \frac{4\pi^2 f^3}{\rho_c} I(f)\,,
\ee
where $\rho_c \equiv 3 H_0^2/8\pi$ is the critical density of the Universe.  

\begin{figure}[htb]
\includegraphics[width=8.5cm]{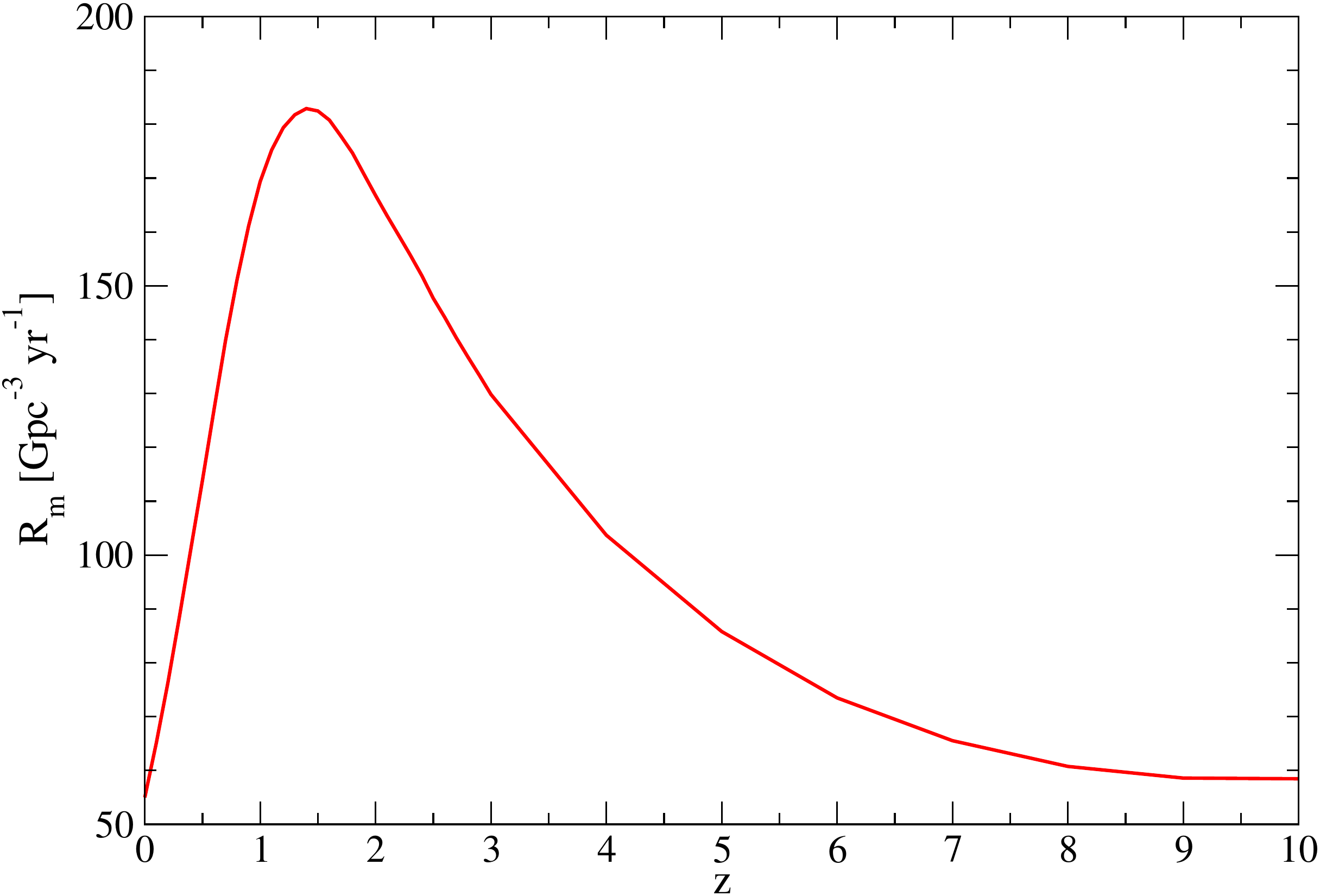}
\caption{Merger rate history of stellar-mass BH binaries $R_m(z)$ with the local rate (at $z=0$) being $R_0 = 55$Gpc$^{-3}$yr$^{-1}$~\cite{TheLIGOScientific:2016pea}.}
\label{fig:Rz}
\end{figure}

We now evaluate  $\Omega_\GW^{(I)} (f)$ generated by stellar-mass BH binaries following~\cite{TheLIGOScientific:2016wyq,Callister:2016ewt}:
\be
\label{eq:Omega-I-integ}
\Omega_\GW^{(I)} (f) = \frac{f}{H_0 \rho_c} \int dz \frac{R_m(z)}{(1+z) \sqrt{\Omega_m(1+z)^3+\Omega_\Lambda}}\frac{dE}{df}\bigg|_{f_s}\,.
\ee
Here $dE/df$ is the emitted GW energy spectrum evaluated at the source frequency $f_s = (1+z) f$ with $f$ representing the observed frequency. As in Sec.~\ref{sec:individual-source}, we use IMRPhenomB waveform model~\cite{Ajith:2009bn} to estimate $dE/df$, which is given e.g. in Eq.~(5) of~\cite{Inayoshi:2016hco}. $R_m(z)$ is the merger rate per comoving volume measured in the source frame (and the $1+z$ factor in the denominator converts this into the detector frame), which we show in Fig.~\ref{fig:Rz}. We assume that the binary BH merger rate follows the convolution between star formation rate~\cite{Vangioni:2014axa} (based on observations of gamma-ray bursts~\cite{Kistler:2013jza}) with metallicity below half the solar metallicity and the probability distribution of lifetimes of binary BHs. We also assume that the local merger rate (at $z=0$) of binary BHs as $R_0 = 55$Gpc$^{-3}$yr$^{-1}$ to fix the overall normalization, which corresponds to the mean value of the event-based estimate with GW150914, GW151226 and LVT151012~\cite{TheLIGOScientific:2016pea}. 

Let us next look at the amount of parity violation in $dE/df$ inside the integral of Eq.~\eqref{eq:Omega-I-integ} in more detail. From~\cite{Phinney:2001di}, one finds 
\be
\label{eq:dEdf}
\frac{dE}{df} \propto \left \langle |\tilde h_R|^2 + |\tilde h_L|^2 \right \rangle_s\,,
\ee
where $\langle \rangle_s$ denotes averaging over source positions and orientations. On the other hand, from Eq.~\eqref{eq:hv}, one finds 
\be
\label{eq:hRL}
\tilde h_{R,L} = \tilde h_{R,L}^\GR (1 + \lambda_{R,L} v)\,.
\ee
Given that no circular polarization exists on average in GR, one can set
\be
\label{eq:h2-GR}
 \left \langle |\tilde h_R^\GR|^2 \right \rangle_s = \left \langle |\tilde h_L^\GR|^2 \right \rangle_s : = \left \langle |\tilde h^\GR|^2 \right \rangle_s\,.
\ee
From Eqs.~\eqref{eq:dEdf} and~\eqref{eq:h2-GR}, one finds
\be
\frac{dE}{df} \propto 2 \left \langle |\tilde h^\GR|^2 \right \rangle_s + \mathcal{O}\left(v^2\right)\,.
\ee
Thus, the parity violation effect in $\Omega_\GW^{(I)}$ enters only at $\mathcal{O}\left(v^2\right)$.

\begin{figure*}[htb]
\includegraphics[width=8.5cm]{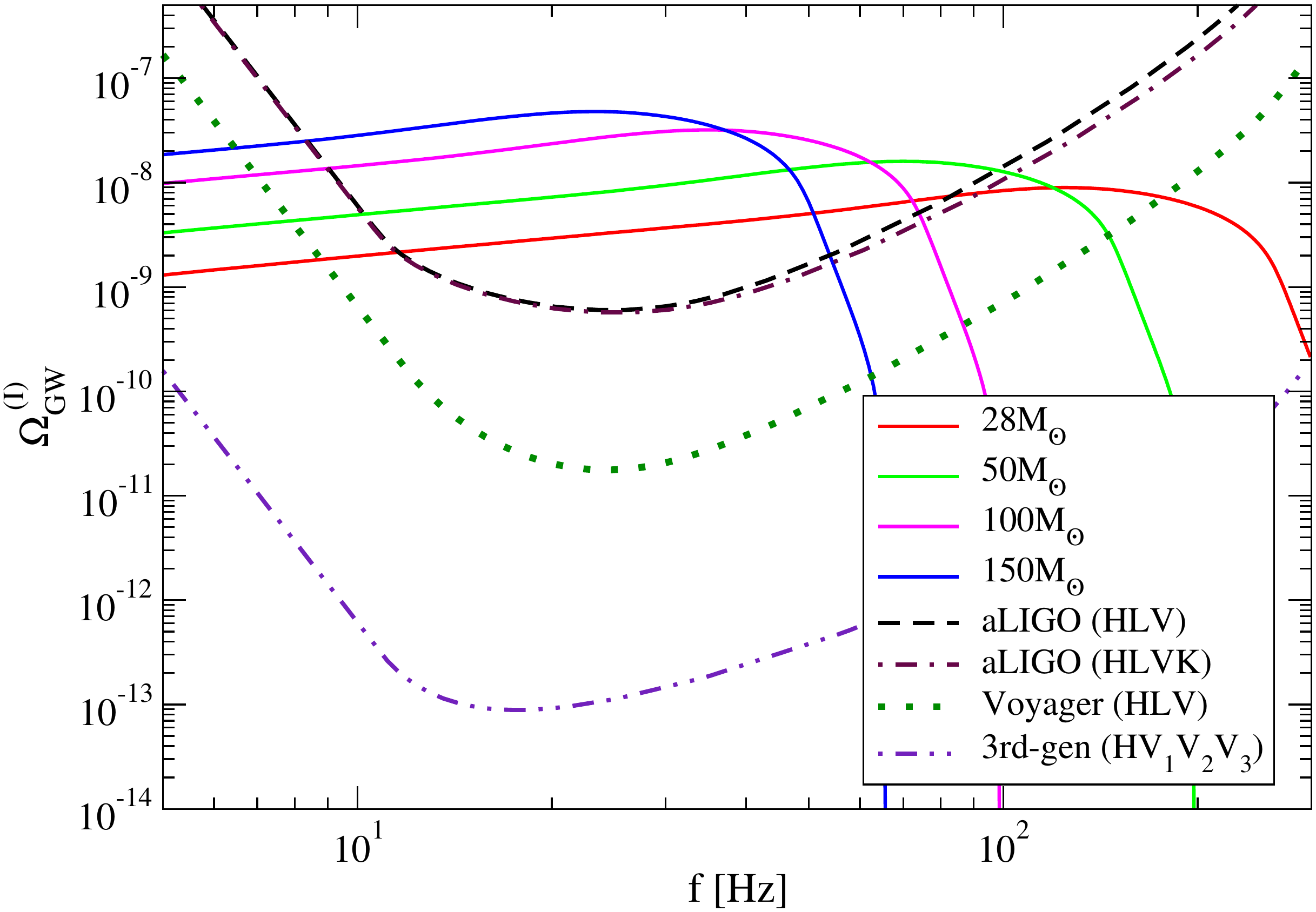}
\includegraphics[width=8.5cm]{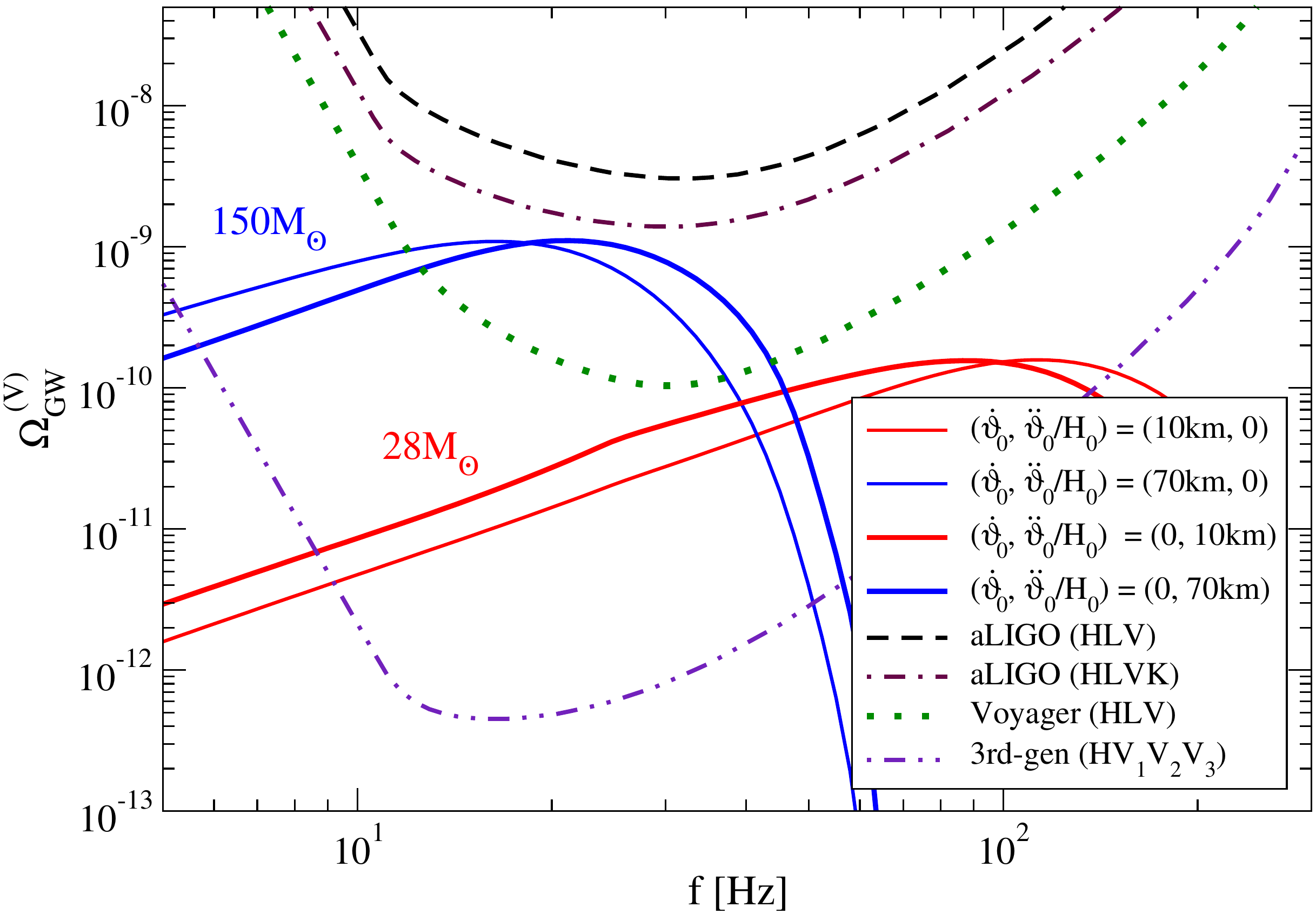}
\caption{(Left) GW energy density spectrum for the I-mode (intensity mode) in GR for various average chirp mass with a local merger rate of $R_0 = 55$Gpc$^{-3}$yr$^{-1}$.  We also present the power-law integrated sensitivity curves~\cite{Thrane:2013oya} for a network of (i) ground-based GW interferometers corresponding to Hanford aLIGO, Livingston aLIGO, Virgo and KAGRA with aLIGO's design sensitivity, (ii) three Voyager detectors at Hanford, Livingston and the Virgo site and (iii) two CE detectors at Hanford and the Virgo site (the Hanford (Virgo) one having one (three) interferometer(s)). If there is a range of frequencies where the GW spectra lies above the sensitivity curves, the SNR is larger than unity. 
(Right) Similar to the left panel but for the V-mode (parity violating mode). We only show the average chirp mass of $28 M_\odot$ (red) and $150M_\odot$ (blue). For each average chirp mass, we consider both $\dot \vartheta = \dot \vartheta_0$ and $\dot \vartheta = \ddot \vartheta_0 t$. Values for $\dot \vartheta_0$ or $\ddot \vartheta_0$ roughly correspond to the maximum ones allowed from the weak CS approximation (see Fig.~\ref{fig:thetadot-thetaddot-pos}). 
}
\label{fig:Omega-GW-I}
\end{figure*}

The left panel of Fig.~\ref{fig:Omega-GW-I} presents $\Omega_\GW^{(I)}$ for various average chirp mass\footnote{The average chirp mass here refers to $\langle M_c^{5/3} \rangle^{3/5}$~\cite{Callister:2016ewt} since $dE/df$ that determines $\Omega_\GW$ (see Eq.~\eqref{eq:Omega-I-integ}) is proportional to $M_c^{5/3}$.} in GR. Observe how the spectrum becomes larger and shifts to a lower frequency as the mass is increased. We here follow~\cite{Callister:2016ewt} and consider the average chirp mass up to $150 M_\odot$. One may think such a chirp mass is too high based on the measured chirp mass of the LVC events. However, these events correspond to nearby sources with $z \leq 0.2$ and the mass distribution for large-$z$ binary BHs may be significantly different from that for low-$z$ sources due to different formation environment, e.g., different metallicity distribution for progenitor stars.

We now define the fractional GW energy density for the V-mode similar to the I-mode one in Eq.~\eqref{eq:Omega-I} as
\be
\Omega_\GW^{(V)} (f) \equiv  \frac{4\pi^2 f^3}{\rho_c} V(f) = \Pi (f)  \Omega_\GW^{(I)} (f)\,,
\ee
where %
\be
\Pi(f) \equiv \frac{V(f)}{I(f)}
\ee
shows the amount of parity violation. As already mentioned, $I$ is proportional to $\langle |\tilde h_R|^2 + |\tilde h_L|^2 \rangle_s$ which can also be seen from Eq.~\eqref{eq:IV}. On the other hand, $V$ is proportional to $\langle |\tilde h_R|^2 - |\tilde h_L|^2 \rangle_s$, which from Eqs.~\eqref{eq:hRL} and~\eqref{eq:h2-GR} becomes
\ba
\left \langle |\tilde h_R|^2 - |\tilde h_L|^2 \right \rangle_s &\propto& 4 \left \langle |\tilde h^\GR|^2 \right \rangle_s  v + \mathcal{O}\left(v^2\right)\,, \nn \\
&\propto & 2 \left \langle |\tilde h_R|^2 + |\tilde h_L|^2 \right \rangle_s v + \mathcal{O}\left(v^2\right)\,. \nn \\
\ea
Using this equation and Eq.~\eqref{eq:Omega-I-integ}, one yields 
\be
\label{eq:Omega-V-integ}
\Omega_\GW^{(V)} (f) = \frac{f}{H_0 \rho_c} \int dz \frac{2 |v| \, R_m(z)}{(1+z) \sqrt{\Omega_m(1+z)^3+\Omega_\Lambda}}\frac{dE}{df}\bigg|_{f_s}\,,
\ee
which is on the order of $\mathcal{O}(v)$. Unlike GWs from individual sources in Sec.~\ref{sec:individual-source}, GWBs can probe the integrated evolution of the scalar field, as $\Omega_\GW^{(V)}$ is given by integrating $v$ over the redshift.

In order to calculate the V-mode stochastic GWBs, one needs to specify the cosmological evolution of the scalar field. We will consider two example cases in CS gravity:

\begin{itemize}

\item[(i)]  $\dot \vartheta = \dot \vartheta_0 + \ddot \vartheta_0 t $

One natural choice of $\dot \vartheta$ is to expand about the current time $t=0$ and keep to linear order in $t$: $\dot \vartheta = \dot \vartheta_0 + \ddot \vartheta_0 t $~\cite{Yunes:2010yf}. One can express this $\dot \vartheta$ in terms of $z$ instead of $t$ by using the relation between these quantities in GR\footnote{Such an analysis corresponds to working in the non-dynamical formulation of CS gravity, in which the Friedmann equation is the same as in GR. This analysis also applies to the dynamical formulation when the CS correction to the Friedmann equation ($\vartheta'^2/2a^2 + V$)~\cite{Dyda:2012rj} is negligible.} given by $t = g(z)/H_0$, where
\allowdisplaybreaks
\ba
\label{eq:g}
g(z) &=&  \int_z^0 \frac{dz}{(1+z)\sqrt{\Omega_m(1+z)^3+\Omega_\Lambda}} \nn \\
&=& \frac{2}{3 \sqrt{\Omega_\Lambda}} \left[ \tanh^{-1}\left(\sqrt{ 1 + (1+z)^3\frac{\Omega_m}{\Omega_\Lambda}  }\right) \right. \nn \\
& & \left. - \tanh^{-1}\left(\sqrt{1 + \frac{\Omega_m}{\Omega_\Lambda} }\right)\right]\,.
\ea
Here $\Omega_m = 0.3$ and $\Omega_\Lambda = 0.7$ are the fractional energy density of dark matter and dark energy. We have assumed a flat universe and neglected the effect of radiation. Using Eqs.~\eqref{eq:Thetadot},~\eqref{eq:v} and~\eqref{eq:g}, we find
\be
\label{eq:v-case1}
v = - \pi f  \left[ \dot \vartheta_0 z + \frac{\ddot \vartheta_0}{H_0} (1+z) g(z) \right]\,.
\ee
%

\item[(ii)]  $V=0$

Instead of Taylor expanding $\dot \vartheta$ about $t=0$, we next consider fixing the scalar field potential $V(\vartheta)$ in dynamical CS gravity. The simplest choice is to set it to zero. In such a case, the scalar field equation is given by $\vartheta'' +2 (a'/a) \vartheta'=0$. Such an equation can easily be solved to yield $\dot \vartheta = \dot \vartheta_0 (1+z)^3$. Substituting this into Eq.~\eqref{eq:Thetadot} and using Eq.~\eqref{eq:v}, one arrives at
\be
v = \pi \dot \vartheta_0 f \left[ 1 - (1+z)^4 \right]\,. 
\ee

\end{itemize}

The right panel of Fig.~\ref{fig:Omega-GW-I} presents $\Omega_\GW^{(V)} (f)$ for the selected average chirp mass in CS gravity with the first example case. We choose the values of  $\dot \vartheta_0$ and $\ddot \vartheta_0$ such that they roughly correspond to the maximum ones satisfying the weak CS approximation, as shown in Fig.~\ref{fig:thetadot-thetaddot-pos}. Notice that such values for $\dot \vartheta_0$ have already been ruled out from binary pulsar observations~\cite{Yunes:2008ua,AliHaimoud:2011bk}, and hence we are using them only for an illustrative purpose.
In addition, the peak frequency of the V-mode spectrum is lower than that of the I-mode spectrum. This is because the former quantity acquires an extra factor of $z$ in the integral in Eq.~\eqref{eq:Omega-V-integ} (originating from the $z$ dependence in $v$), which puts more weight on BH binaries with larger redshifts and brings the overall redshifted frequency lower.

\subsection{Data Analysis Formalism}

In this section, we review the formalism to detect circular polarizations using ground-based GW interferometers developed in~\cite{Seto:2007tn,Seto:2008sr}. 
GWBs can be detected by cross-correlating signals from two or more detectors. For example, the cross-correlated SNR of signals from the $a$th and $b$th detectors with a coincident observation time $T$ is given by
\be
\rho = \frac{3H_0^2}{10\pi^2} \sqrt{2T} \left[ \int^\infty_0 df \frac{\Omega_\GW^{(I)}{}^2 (\gamma_{I,ab} + \gamma_{V,ab} \Pi)^2}{f^6 S_{n,a} S_{n,b}} \right]^{1/2}\,,
\ee
where $S_{n,A}$ is the noise spectral density of the $A$th detector. $\gamma_{I,ab}$ and $\gamma_{V,ab}$ are the overlap reduction functions for the I and V mode~\cite{Allen:1997ad} given by
\ba
\gamma_{I,ab} &\equiv& \frac{5}{8\pi} \int d\Omega\, e^{2\pi i f n\cdot \Delta \vec x} (F_{+,a}F_{+,b}+F_{\times,a}F_{\times,b})\,, \\
\gamma_{V,ab} &\equiv& \frac{5}{8\pi} i \int d\Omega\, e^{2\pi i f n\cdot \Delta \vec x} (F_{+,a}F_{\times,b}-F_{\times,a}F_{+,b})\,, \nn \\
\ea
with $\Delta \vec x \equiv \vec x_a - \vec x_b$.
The SNR becomes smaller than that with two detectors located at the same site by the overlap reduction functions due to the time delay between the two detectors and the misalignment of the detector arms. 

One can separate out the I-mode and V-mode by correlating signals from three or more detectors. Let us assume that one has $n_t$ signal pairs with an identical noise sensitivity. The SNR of each mode is given by~\cite{Seto:2008sr}
\allowdisplaybreaks
\ba
\label{eq:rho-I}
\rho_I &=& \frac{3H_0^2}{10\pi^2} \sqrt{2T} \left[ \int^\infty_0 df\, \frac{\Omega_\GW^{(I)}{}^2 \,\bar \gamma_{I}^2}{f^6 S_{n}^2} \right]^{1/2}\,, \\
\label{eq:rho-V}
\rho_V &=& \frac{3H_0^2}{10\pi^2} \sqrt{2T} \left[ \int^\infty_0 df\, \frac{\Omega_\GW^{(I)}{}^2\, \Pi^2\, \bar \gamma_{V}^2}{f^6 S_{n}^2} \right]^{1/2}\,, 
\ea
where $\bar \gamma_{I}$ and $\bar \gamma_{V}$ are the effective compiled overlap reduction functions given by
\ba
\label{eq:gamma-I}
\bar \gamma_{I} &\equiv& \left( \frac{\sum_i^{n_t} \gamma_{I,i}^2 \sum_i^{n_t} \gamma_{V,i}^2- (\sum_i^{n_t} \gamma_{I,i} \gamma_{V,i})^2}{\sum_i^{n_t} \gamma_{V,i}^2} \right)^{1/2}\,, \\
\label{eq:gamma-V}
\bar \gamma_{V} &\equiv& \left( \frac{\sum_i^{n_t} \gamma_{I,i}^2 \sum_i^{n_t} \gamma_{V,i}^2- (\sum_i^{n_t} \gamma_{I,i} \gamma_{V,i})^2}{\sum_i^{n_t} \gamma_{I,i}^2} \right)^{1/2}\,, 
\ea
with $\gamma_{I,i}$ and $\gamma_{V,i}$ representing the overlap reduction function of the $i$th pair. Notice that SNRs for stochastic GWBs scale with $1/S_n$, while those for GWs from individual sources scale with $1/S_n^{1/2}$. Thus, bounds on parity violation with GWBs are more sensitive to the improvement in the detector sensitivity than those from individual GW sources, as we will see in more detail later.

\begin{figure}[htb]
\includegraphics[width=8.5cm]{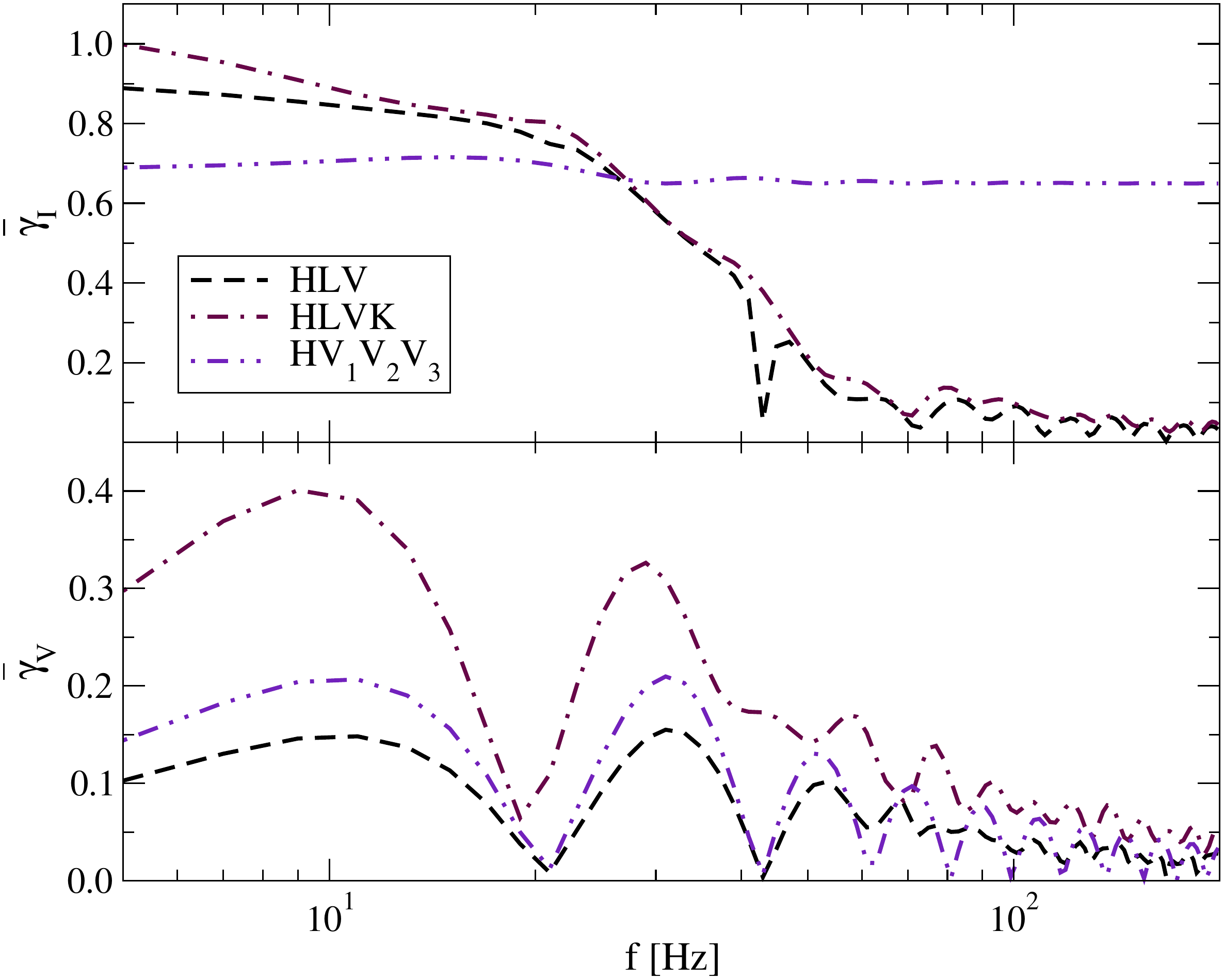}
\caption{Effective compiled overlap reduction functions for the I-mode in Eq.~\eqref{eq:gamma-I} (top) and the V-mode in Eq.~\eqref{eq:gamma-V} (bottom) with a network of three (Hanford, Livingston and Virgo), four (Hanford, Livingston, Virgo and KAGRA) and four (one at Hanford site and three at the Virgo site) GW interferometers. 
}
\label{fig:Gamma-eff}
\end{figure}

Figure~\ref{fig:Gamma-eff} illustrates $\bar \gamma_{I}$ and $\bar \gamma_{V}$ for a network of three and four second-generation GW interferometers. Observe that the overlap reduction functions for the I-mode drops rapidly at $c/R_\oplus \sim 47$Hz (where $R_\oplus$ is the Earth radius). Observe also that the fourth detector does not contribute much for the I-mode, while it improves the sensitivity of the V-mode by a factor of $\sim 2$.  Such a finding is consistent with Table IV in~\cite{Seto:2008sr}. In general, the V-mode sensitivity is much smaller than the I-mode sensitivity. 

We also consider two third-generation GW detectors (both having a sensitivity comparable to the CE one) one at Hanford and the other at the Virgo site. We assume that the Hanford detector (H) has the traditional ``L-shape" configuration as CE while the Virgo-site detector ($V_{1,2,3}$) has the same ``triple-interferometer" configuration as ET.
For simplicity, we assume that the orientation of the bisector of the first interferometer $V_1$ is same as that of Virgo. The remaining two interferometers $V_2$ and $V_3$ are obtained by rotating $V_1$ by $2\pi/3$ and $4\pi/3$ respectively in the detector plane.

We present in Fig.~\ref{fig:Gamma-eff} $\bar \gamma_{I}$ and $\bar \gamma_{V}$ for four CE interferometers.  Observe first that $\bar \gamma_{I}$ is almost constant. This is because one detector at the Virgo site forms three independent interferometers and the overlap reduction function between such interferometers does not drop until $f \sim 1$kHz~\cite{Regimbau:2012ir}. On the other hand, observe that $\bar \gamma_{V}$ for CE is more similar to that for HLV than HLVK. This is because if the two interferometers are coplanar, their correlation is insensitive to circular polarizations. Thus, V$_1$V$_2$, V$_1$V$_3$ and V$_2$V$_3$ do not contribute and only three pairs are sensitive to such polarizations (while there are three (six) pairs for HLV (HLVK)). Since the detector pairs for circular polarizations are always constructed from one interferometer at Hanford and one at the Virgo site, $\bar \gamma_{V}$ for HV$_1$V$_2$V$_3$ shows a similar behavior as that for HLV and HLVK.

Using Eqs.~\eqref{eq:rho-I} and~\eqref{eq:rho-V}, we present in Fig.~\ref{fig:Omega-GW-I} power-law integrated noise curves~\cite{Thrane:2013oya} for a network of second generation ground-based detectors with each detector having the sensitivity equivalent to aLIGO's design sensitivity. We also present such noise curves for a network of three Voyager detectors at Hanford, Livingston and the Virgo site, and a network of two CE detectors that consists of four independent interferometers as mentioned in the previous paragraph. If any part of the GW spectrum lies above a particular noise curve, the SNR is above unity. Observe that a network of three Voyager (two CE) detectors is more sensitive than that for a network of three second-generation detectors by $\sim 2$ ($\sim 4$) orders of magnitude. This reflects the ratio between $S_n$ for Voyager (CE) and aLIGO (see Fig.~\ref{fig:noise}).

\subsection{Results}

We now explain how strongly one can constrain gravitational parity violation with stochastic GWBs from BH binaries assuming that the signal is consistent with GR.  We set the detection threshold of the V-mode as 5. For a flat GW energy density spectrum, the threshold SNR is given by~\cite{Allen:1997ad}
\be
\rho_\mrm{thr} = \sqrt{2} \left[ \mrm{erfc}^{-1} (2 P_f) - \mrm{erfc}^{-1} (2 P_d) \right]\,,
\ee
where $P_f$ and $P_d$ are the false alarm rate and the detection rate respectively, while $\mrm{erfc}(x)$ is the complementary error function. Thus, $\rho_\mrm{thr} = 5$ corresponds to e.g.~$P_f = 3.8 \times 10^{-3}$ and $P_d = 0.99$.

One important difference between the GWB bound on gravitational parity violation and bounds obtained by other methods, such as GWs from individual sources, solar system experiments and binary pulsar observations, is that the GWB bound depends on the cosmological history of gravitational parity violation. This is because the GWB spectrum is obtained by integrating the parity violation effect over the redshift (see Eq.~\eqref{eq:Omega-V-integ}), while bounds from nearby sources can only probe parity violation at $z \ll 1$.

In the discussion below, we derive specific bounds on CS gravity. We first consider an ideal situation where all the source parameters (except for the CS scalar field and the BH merger rate history) are known. We next consider a more realistic situation where one does not have such information beforehand.
We end this section by comparing our analysis with the Bayesian parameter estimation study in~\cite{Crowder:2012ik} for the case where the parity violation $v$ does not depend on $f$ nor $z$.

\subsubsection{CS Gravity: Ideal Case}

We begin with the ideal situation. Regarding the scalar field evolution models, we mostly focus on the first model in Sec.~\ref{sec:GW-spectrum} ($\dot \vartheta = \dot \vartheta_0 + \ddot \vartheta_0 t$), and will comment on how the bounds change if one considers the second model ($V=0$). 

Figure~\ref{fig:thetadot-thetaddot-pos} presents the upper bounds on positive $\dot \vartheta_0$ and $\ddot \vartheta_0/H_0$ for a network of various detectors with the average chirp mass of $28M_\odot$ (top) and $150M_\odot$ (bottom). The bounds satisfy the weak CS approximation if they lie within the blue shaded region. Notice first that bounds associated with Voyager-type detectors may marginally satisfy the weak CS condition if the average chirp mass is large, while one needs a network of CE-type detectors to place meaningful bounds on CS gravity if the average chirp mass is small. Notice also that GW bounds are complementary to existing bounds from solar system experiments and binary pulsar observations as they are only sensitive to  $\dot \vartheta_0$. 

Let us study the behavior of projected constrained parameter regions in Fig.~\ref{fig:thetadot-thetaddot-pos} in more detail. One sees that such regions are unbounded and when $\dot \vartheta_0$ and $\ddot \vartheta_0/H_0$ become larger, the allowed ranges become narrower around the relation $\ddot \vartheta_0/H_0 = \dot \vartheta_0$. This is because GWB observations place bounds on a certain combination of $\dot \vartheta_0$ and $\ddot \vartheta_0/H_0$ that roughly corresponds to that in Eq.~\eqref{eq:v-case1}, namely $\dot \vartheta_0 z + (\ddot \vartheta_0/H_0) (1+z) g(z)$. When $\dot \vartheta_0$ and $\ddot \vartheta_0/H_0$ are large, the two terms in the combination needs to cancel with each other almost exactly so that the combination becomes a small number that satisfies the observational bounds. 
 Interestingly, the ratio between the coefficients of $\dot \vartheta_0$ and $(\ddot \vartheta_0/H_0)$ in this combination is $\sim -1$ irrespective of the value of $z$. Thus the relation between $\dot \vartheta_0$ and $\ddot \vartheta_0/H_0$ needs to be $\ddot \vartheta_0/H_0 \approx \dot \vartheta_0$ in order to realize such a kind of cancellation.

\begin{figure}[htb]
\includegraphics[width=8.5cm]{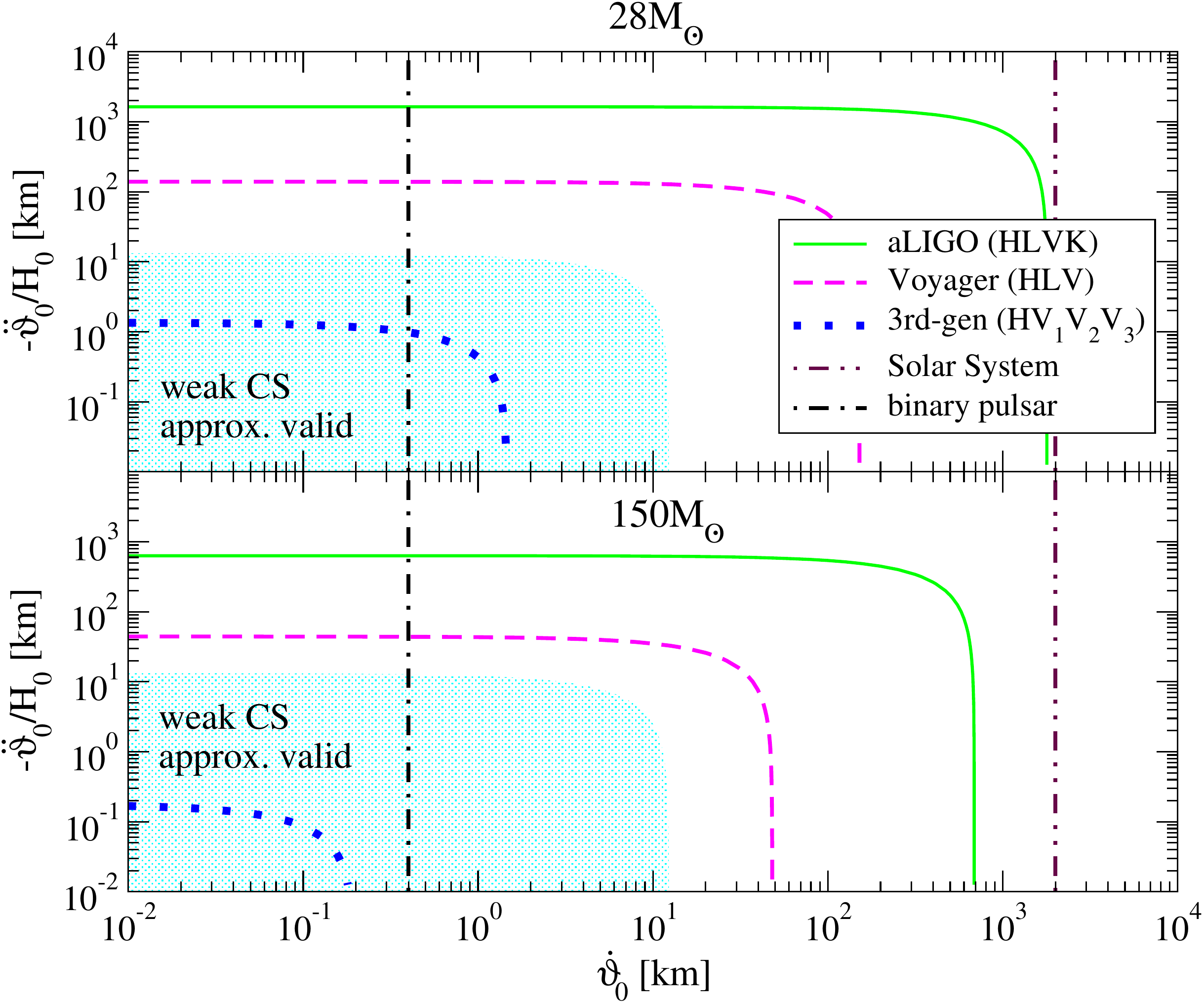}
\caption{Similar to Fig.~\ref{fig:thetadot-thetaddot-pos} but for negative $\ddot \vartheta_0$.
}
\label{fig:thetadot-thetaddot-neg2}
\end{figure}

Figure~\ref{fig:thetadot-thetaddot-pos} shows the bounds on positive $\dot \vartheta_0$ and $\ddot \vartheta_0/H_0$, and the cases with negative $\dot \vartheta_0$ and/or negative $\ddot \vartheta_0/H_0$  need to be addressed separately. When $\dot \vartheta_0$ and $\ddot \vartheta_0/H_0$ are both negative, one again finds bounds in Fig.~\ref{fig:thetadot-thetaddot-pos} but replacing $(\dot \vartheta_0, \ddot \vartheta_0/H_0) \to (-\dot \vartheta_0, -\ddot \vartheta_0/H_0)$.  On the other hand, Fig.~\ref{fig:thetadot-thetaddot-neg2} presents the bounds on positive $\dot \vartheta_0$ and negative $\ddot \vartheta_0/H_0$. In such a case, the cancellation between the two terms mentioned in the previous paragraph never occurs and the constrained regions are bounded from above. When $\dot \vartheta_0$ is negative and $\ddot \vartheta_0/H_0$ is positive instead, the bounds would be the same as in Fig.~\ref{fig:thetadot-thetaddot-neg2} but replacing $(\dot \vartheta_0, - \ddot \vartheta_0/H_0) \to (-\dot \vartheta_0, \ddot \vartheta_0/H_0)$.

\begin{figure}[htb]
\includegraphics[width=8.5cm]{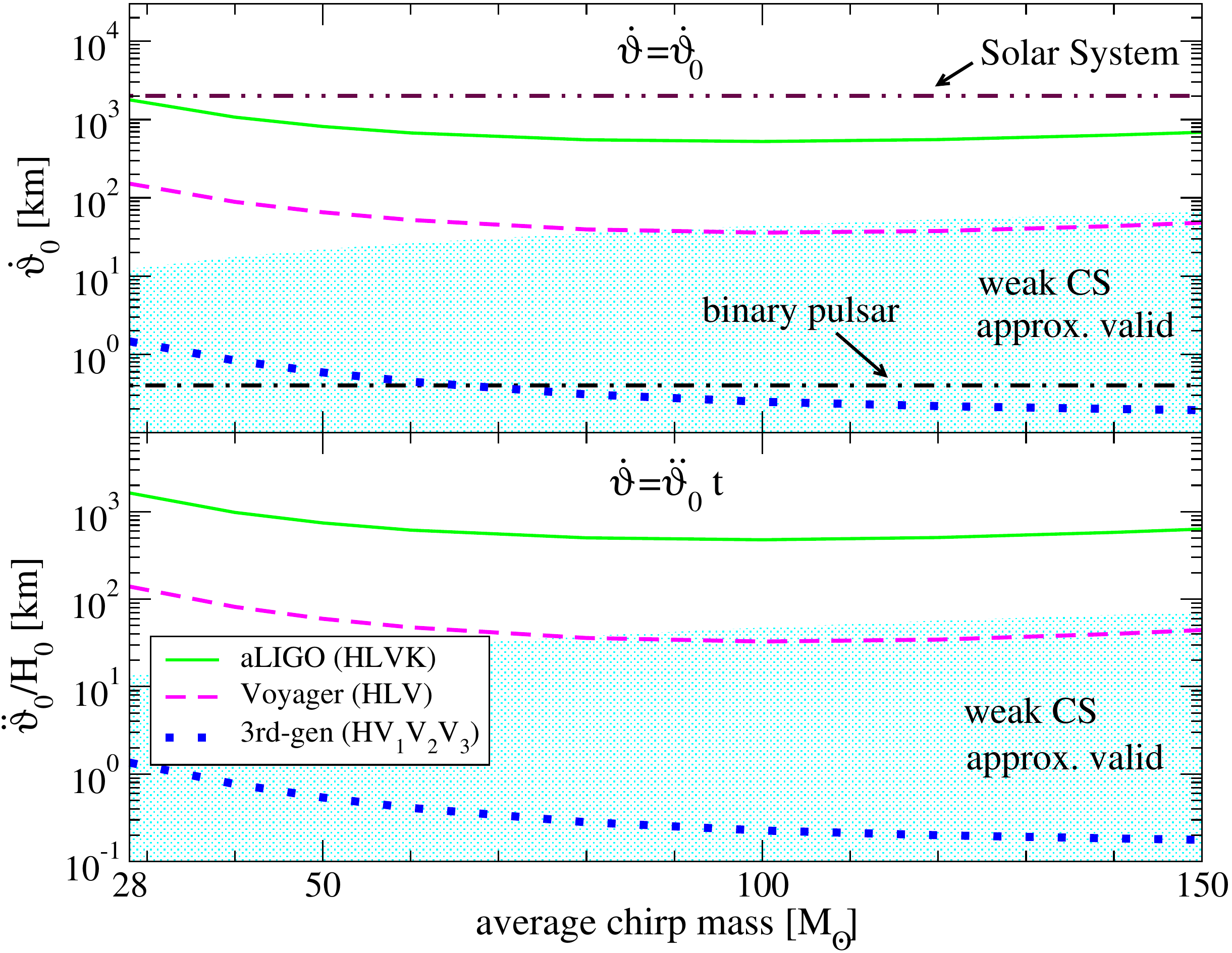}
\caption{(Top) Upper bounds on the evolution of the scalar field $\dot \vartheta = \dot \vartheta_0$ in CS gravity from stochastic GWBs of stellar-mass BH binaries as a function of the average chirp mass. We assume using a network of four second-generation ground-based GW interferometers (green solid), three Voyager-type detectors (magenta dashed) and two CE-type detectors (blue dotted). The bounds are derived within the weak CS approximation and are valid only in the blue shaded region. We also show the bound from the solar system (LAGEOS) experiment~\cite{Smith:2007jm} and binary pulsar observations~\cite{Yunes:2008ua,AliHaimoud:2011bk}. Observe that CE bounds always satisfy the weak CS approximation and can be comparable to or even slightly stronger than the binary pulsar one. 
(Bottom) Similar to the top panel but for the upper bounds on $\ddot \vartheta = \ddot \vartheta_0$, where we consider the case $\dot \vartheta= \ddot \vartheta_0\, t$. Solar system experiments and binary pulsar observations are not sensitive to $\ddot \vartheta_0$. 
}
\label{fig:theta-dot}
\end{figure}

We now look at how the bounds on $\dot \vartheta_0$ and $\ddot \vartheta_0$ depend on the average chirp mass of  BH binaries in more detail. 
The green solid curve in the top panel of Fig.~\ref{fig:theta-dot} shows the bounds on $\dot \vartheta_0$ for the case that $\dot \vartheta = \dot \vartheta_0$. Notice that such constraint becomes most stringent  in the intermediate (average) chirp mass regime. 
The blue shaded region shows the parameter space in which the weak CS approximation is satisfied. The boundary of this region is given by setting the frequency in the approximation in Eq.~\eqref{eq:weak-CS} or~\eqref{eq:weak-CS2} to be the termination frequency of the IMRPhenomB waveform while we choose the maximum redshift to be $z=10$~\footnote{This choice of $f$ and $z$ for finding the boundary of the weak CS approximation region is used also for the $\dot \vartheta = \ddot \vartheta_0 t$ and $V=0$ cases later.}. 
Notice also that the bound associated with aLIGO-type detectors lies outside such ``weak-CS" regime, so that its validity remains questionable. For comparison, we show the bounds from solar system experiments and binary pulsar observations. 
Although Fig.~\ref{fig:theta-dot} assumes using a network of four ground-based detectors (HLVK), we have also checked that the bounds become worse by a factor of a few if one considers instead a network of three ground-based detectors (HLV). 

We next study the improvement of constraints on CS scalar field with advanced detectors.
The magenta dashed curve in the top panel of Fig.~\ref{fig:theta-dot} shows the bound on $\dot \vartheta_0$ using a network of three Voyager detectors. Such a bound is better than those associated with second-generation by roughly one order of magnitude. It marginally satisfies the weak CS approximation for relatively large average chirp masses ($\ge 90 M_{\odot}$),
and it is always weaker than the binary pulsar constraint. 
The blue dotted curve presents the bound with a network of two CE detectors, which universally satisfies the weak CS condition in the mass range considered here. This bound is stronger than the second-generation detector bound by $\sim 3$ orders of magnitude, and becomes comparable to or even slightly stronger than the binary pulsar bound for large average chirp masses ($\ge 65 M_\odot$)\footnote{Though it is likely that the binary pulsar bounds will improve by the time CE detectors operate.}. Such an enhancement in CE from the second-generation case is much larger than that by using GWs from individual sources, which improves the constraint by only  $\sim 1$ order of magnitude (see Table~\ref{table:bounds}). Thus, bounds on gravitational parity violation with stochastic GWBs are more sensitive to the detector sensitivity improvement than those with GWs from individual sources.

The bottom panel of Fig.~\ref{fig:theta-dot} is similar to the top panel, except that it presents the bounds on $\ddot \vartheta_0$ for the model $\dot \vartheta = \ddot \vartheta_0 t$. Bounds from solar system experiments and binary pulsar observations are not sensitive to $\ddot \vartheta_0$. 
Similar to the top panel, we find that that the bound with a network of four second-generation detectors does not satisfy the weak CS approximation, while that with three Voyager detectors marginally satisfies the approximation for large average chirp masses ($\ge 70 M_\odot$). The bound obtained by using a network of two CE detectors are stronger than the one with second-generation detectors by more than $\sim 3$ orders of magnitude and it is always consistent with the weak CS approximation.

 \begin{figure}[htb]
\includegraphics[width=8.5cm]{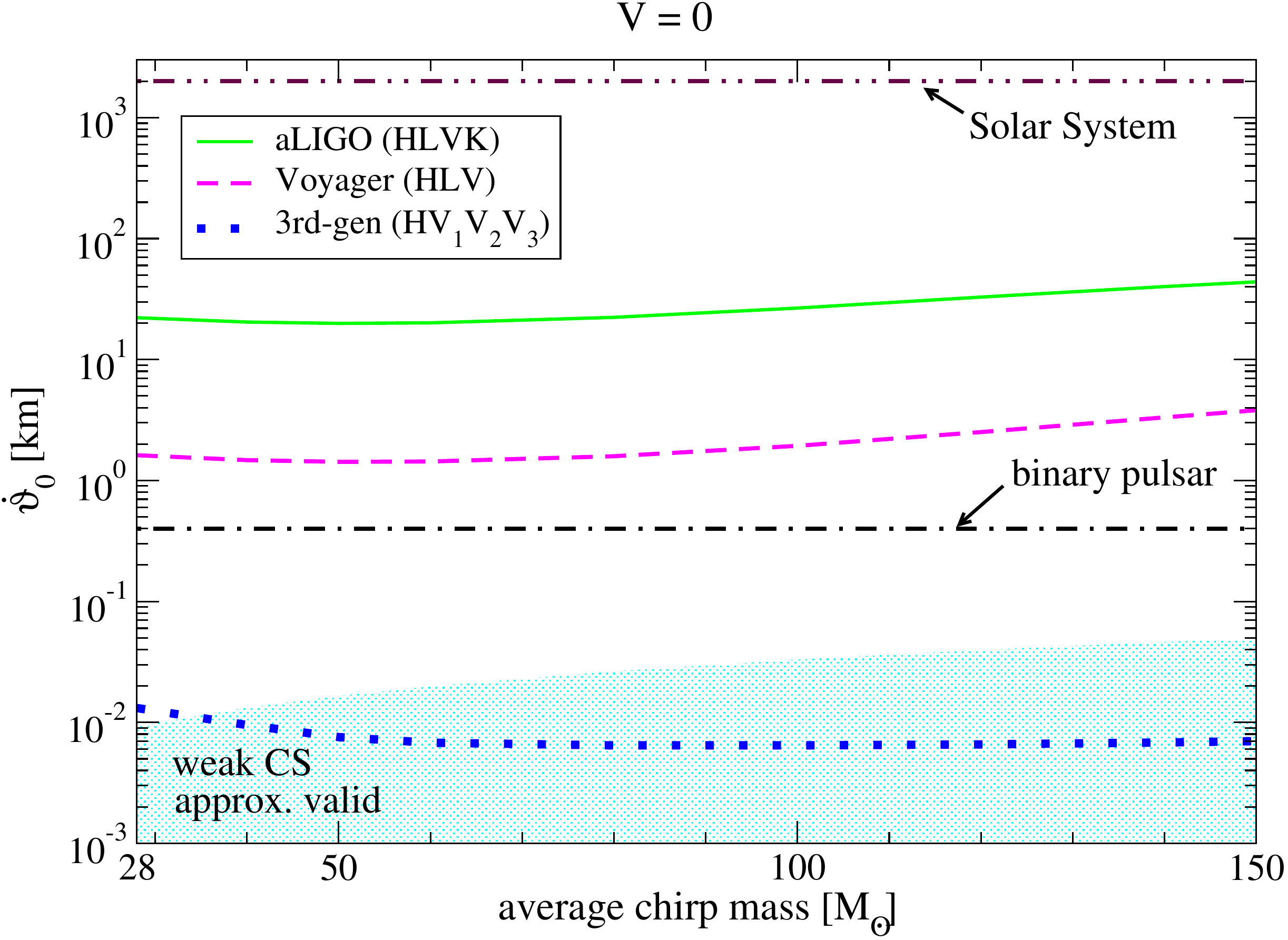}
\caption{Similar to the top panel of Fig.~\ref{fig:theta-dot} but for the $V=0$ case. 
}
\label{fig:theta-dot-V0-2-new}
\end{figure}

We now consider a model with $V=0$ that is different from the one assuming $\dot \vartheta = \dot \vartheta_0 + \ddot \vartheta_0 t$.
Figure~\ref{fig:theta-dot-V0-2-new} presents bounds on $\dot \vartheta_0$ for the $V=0$ model as a function of the average chirp mass for BH binaries.
We find that the weak CS condition is much more difficult to be met than the $\dot \vartheta = \dot \vartheta_0$ case and both bounds from a network of second-generation detectors and Voyager-type detectors do not satisfy the condition. On the other hand, bounds obtained from a network of two CE detectors mostly satisfy the condition, and such bounds are stronger than the binary pulsar bound by more than one order of magnitude\footnote{Strictly speaking, one cannot directly compare the GW bounds with the solar system or binary pulsar bounds as the latter two were derived under the assumption that $\ddot \vartheta = 0$.}.

\subsubsection{CS Gravity: More Realistic Case}

Up until now, we have been considering an ideal situation where one has a perfect knowledge of the merger rate history and average BH parameters (like masses and spins) that produce stochastic GWBs, so that one can immediately derive bounds on parity violation from a non-detection of the V-mode GW spectrum. In reality, one does not have such precise information beforehand. In this subsection, we will discuss how one can obtain approximate bounds in practice. 

The key is to use $\Omega_\GW^{(I)}$, which should be measured separately. Since the binary BH merger rate that follows the observed star formation rate has a peak around $z_\mrm{peak} \approx 1.5$ (see Fig.~\ref{fig:Rz}), the dominant contribution in the $z$ integral in Eq.~\eqref{eq:Omega-V-integ} should come around this peak $z$. Thus, one can approximate this equation as
\ba
\label{eq:Omega-V-integ-approx}
\Omega_\GW^{(V)} (f) & \approx &  2  |v(z_\mrm{peak})|  \frac{f}{H_0 \rho_c} \nn \\
&& \times  \int dz \frac{R_m(z)}{(1+z) \sqrt{\Omega_m(1+z)^3+\Omega_\Lambda}}\frac{dE}{df}\bigg|_{f_s}\,, \nn \\
&\approx & 2  |v(z_\mrm{peak})| \, \Omega_\GW^{(I)} (f)\,, 
\ea
where we used Eq.~\eqref{eq:Omega-I-integ} in the last equality. Notice that $\Omega_\GW^{(V)}$ is now simply given by a directly measurable quantity $\Omega_\GW^{(I)}$ and $v$ at the peak frequency $z_\mrm{peak}$. Thus, the only a priori information one needs is $z_\mrm{peak}$, which again should be around 1.5 as long as the binary BH merger rate history follows that of the star formation rate. 
 
 \begin{figure}[htb]
\includegraphics[width=8.5cm]{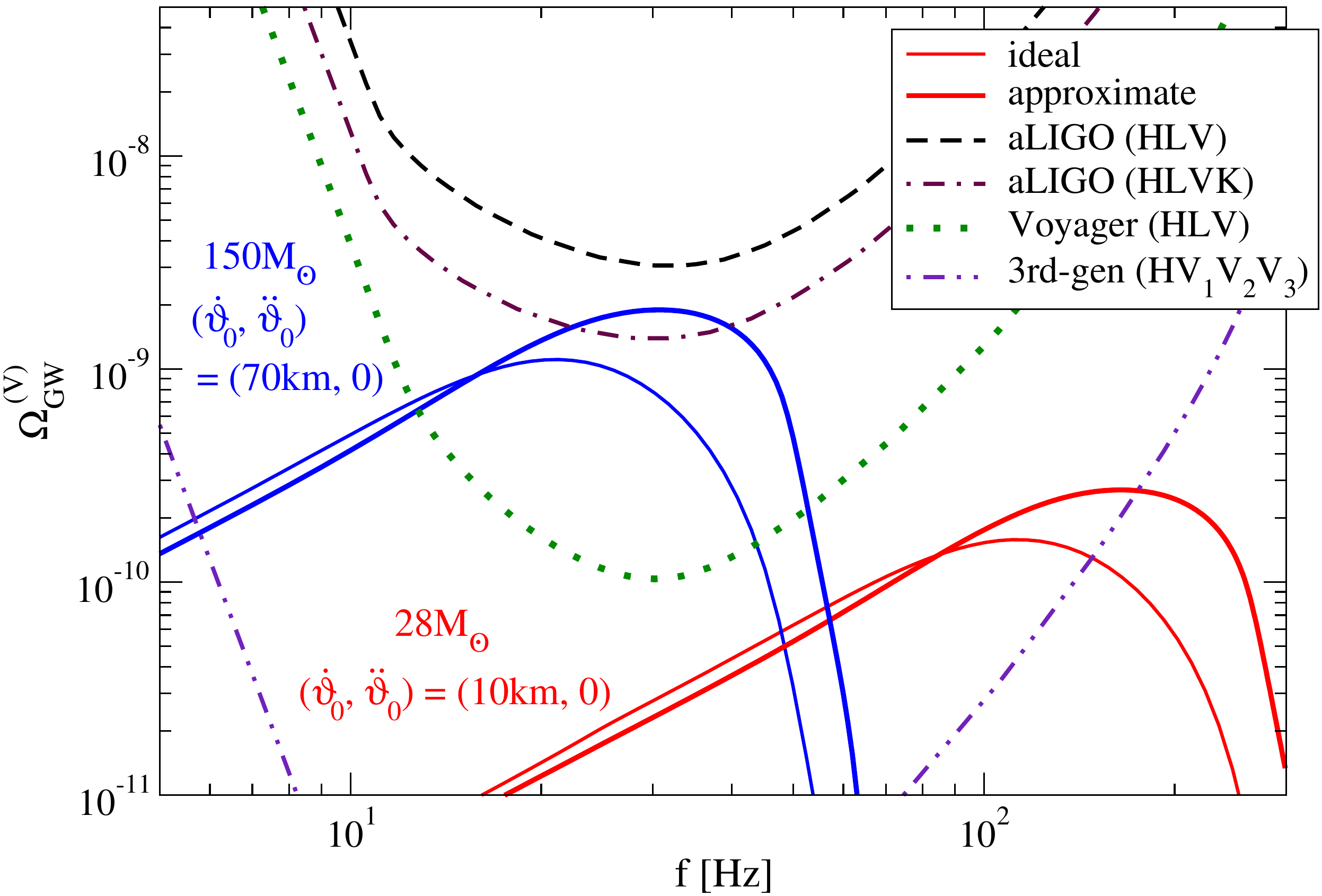}
\caption{Comparison of the V-mode spectrum $\Omega_\GW^{(V)}$ for the $\dot \vartheta = \dot \vartheta_0$ case in the ideal situation where the merger history and other binary BH parameters are known exactly (Eq.~\eqref{eq:Omega-V-integ}) and in the more realistic situation (Eq.~\eqref{eq:Omega-V-integ-approx}). In the latter case, $\Omega_\GW^{(V)}$ is approximated from the knowledge of $\Omega_\GW^{(I)}$ and the peak redshift of the BH merger rate. Observe that the approximation becomes more accurate on the low frequency part of the spectrum, while it overestimates the peak values. 
}
\label{fig:Omega-GW-V-approx}
\end{figure}

Figure~\ref{fig:Omega-GW-V-approx} compares $\Omega_\GW^{(V)}$ for the ideal and realistic situations for the $\dot \vartheta = \dot \vartheta_0$ case. Observe that the latter approximates the former relatively accurately in the low frequency regime. On the other hand, the approximation  overestimates $\Omega_\GW^{(V)}$ around its peak for each spectrum. This is because the peak frequency of the V-mode GW energy density spectrum is lower than that of the I-mode spectrum, as already mentioned earlier (see Fig.~\ref{fig:Omega-GW-I}), and the approximate spectrum of the V-mode (obtained from Eq.~\eqref{eq:Omega-V-integ-approx}) follows the frequency dependence of the I-mode. Therefore the peak frequency of the approximate V-mode spectrum appears higher than the actual one.

\begin{figure}[htb]
\includegraphics[width=8.5cm]{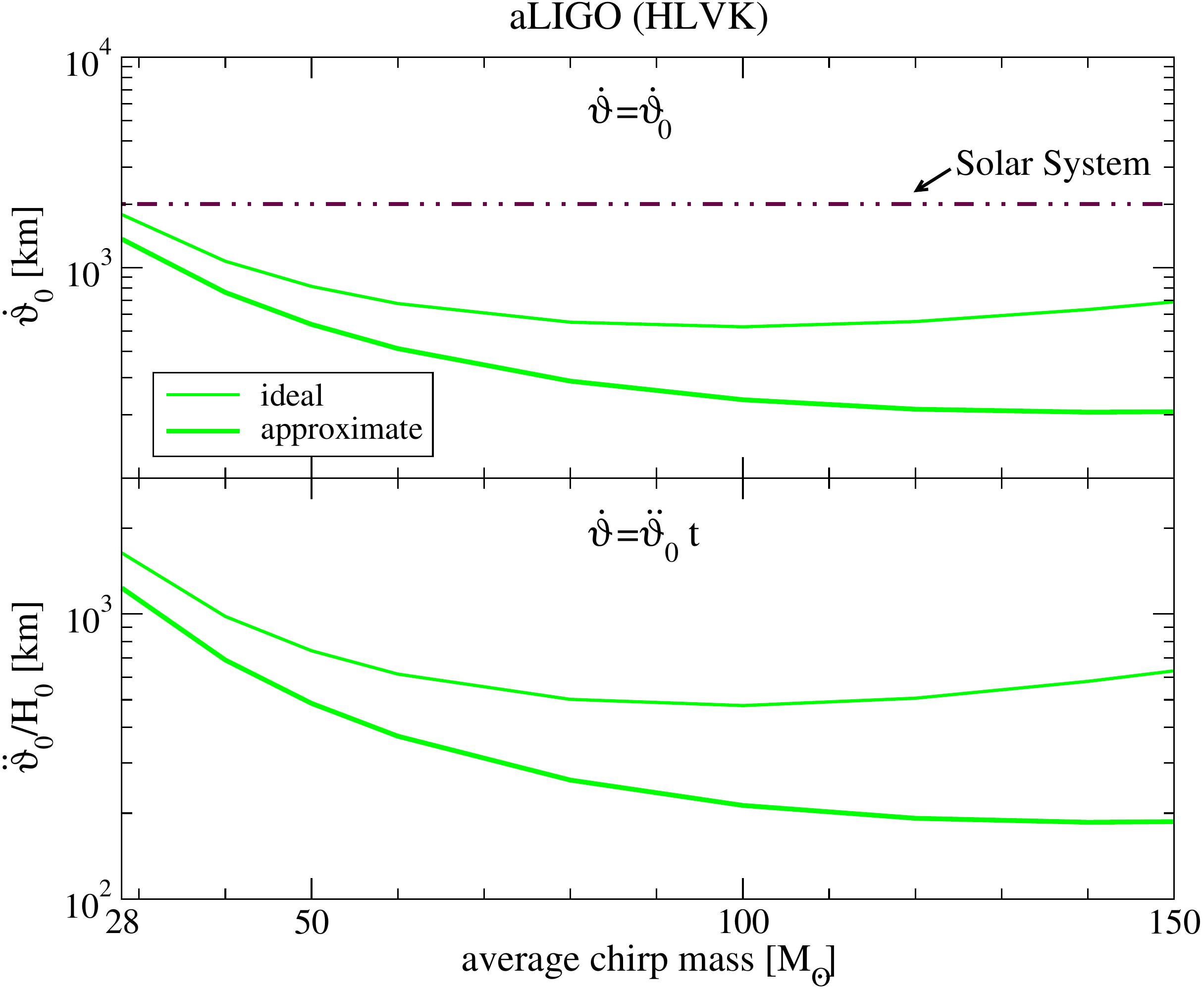}
\caption{Similar to the bounds on $\Omega_\GW^{(V)}$ using a network of four second-generation GW detectors, but we here compare ideal (thin) and more realistic  (thick) situations. Observe that one can still find bounds that are correct as an order of magnitude estimate even if the merger history and GR parameters are not completely known.
}
\label{fig:theta-dot-history}
\end{figure}

Figure~\ref{fig:theta-dot-history} compares the bounds on $\dot \vartheta_0$ and $\ddot \vartheta_0$ with a network of four second-generation GW detectors using the approximate $\Omega_\GW^{(V)}$ with those for the ideal case. One sees that the approximate bounds are quite accurate for the lower average chirp mass, where the peak of the GW spectrum lies outside of the detector frequency band. On the other hand, when the average chirp mass is larger, the peak lies in the detector's sensitive frequency range (see Fig.~\ref{fig:Omega-GW-V-approx}), and thus the approximated bounds should only be taken as an order of magnitude estimate. One finds a similar behaviour for bounds with Voyager or CE.

\subsubsection{Comparison with a Bayesian Parameter Estimation Analysis}

In Sec.~\ref{sec:individual-source}, we have derived bounds on CS gravity with GWs from individual sources with a parameter estimation analysis using Fisher method, while in this section, we have so far derived bounds from stochastic GWBs with a model selection analysis between the detection and non-detection hypotheses of the V-mode GWs. One immediate question is that how would the results change if one uses a parameter estimation analysis instead. To address this question, we compare our analysis with a Bayesian parameter estimation study in~\cite{Crowder:2012ik} based on~\cite{Mandic:2012pj}, in which the authors derived bounds on a parity violation parameter $v$ that does not depend on $f$ nor $z$ and assume that $\Omega_\GW^{(I)}$ is given by a single power law of the form $\Omega_\GW^{(I)} = \Omega_\alpha (f/f_\mrm{ref})^{\alpha}$, where $\Omega_\alpha$ is the overall magnitude of the spectrum, $f_\mrm{ref}$ is the reference frequency while $\alpha$ is the power-law index, which becomes $\alpha = 2/3$ for GWBs from compact binary inspirals~\cite{Romano:2016dpx}. In such a case with a constant $v$, $\Omega_\GW^{(I)}$ and $\Omega_\GW^{(V)}$ have  identical frequency dependence.  

\begin{figure}[htb]
\includegraphics[width=8.5cm]{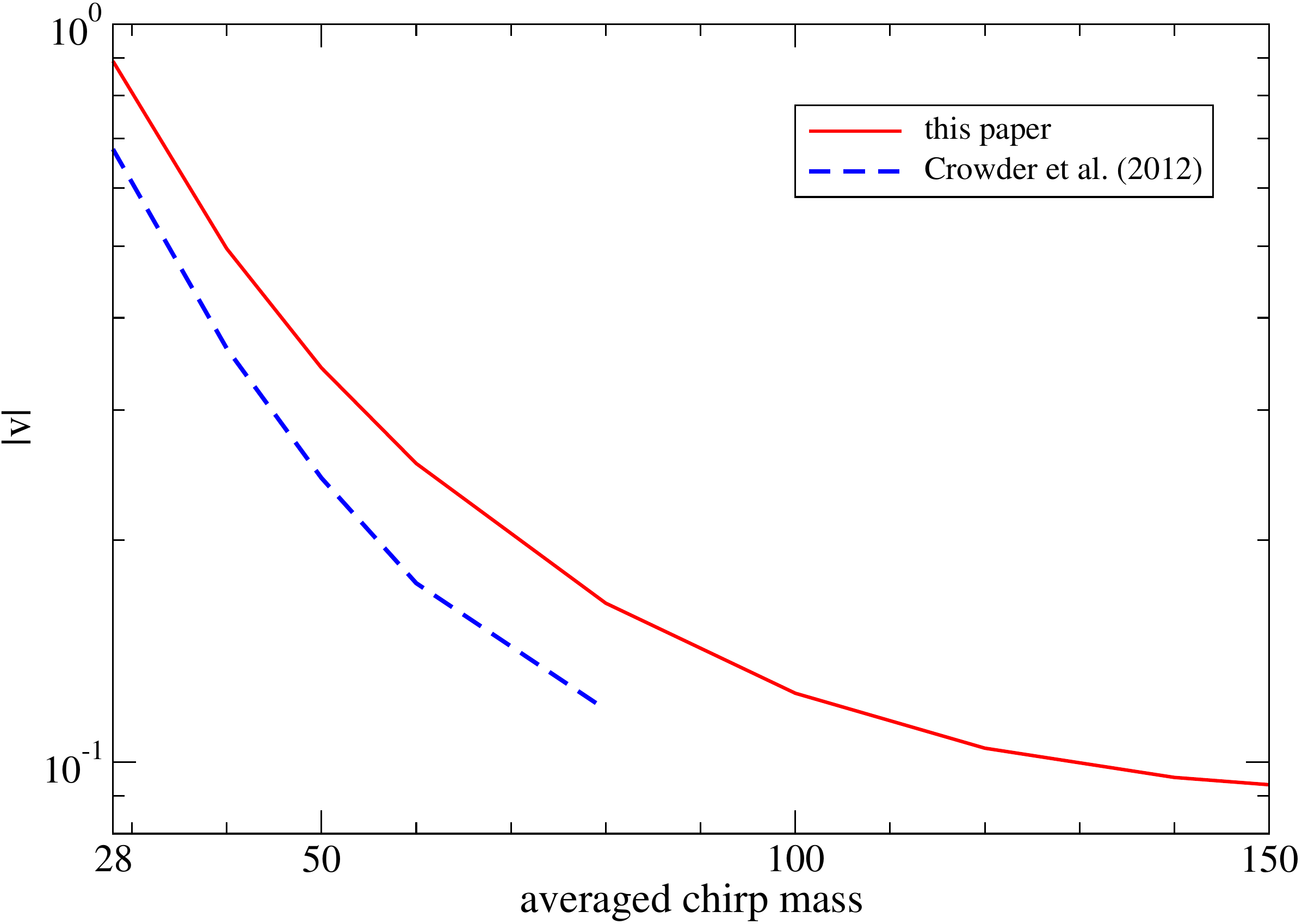}
\caption{Upper bounds on the constant $v$ in this paper (red solid) and in Crowder \textit{et al}.~\cite{Crowder:2012ik} (blue dashed) as a function of the average chirp mass for BH binaries that create stochastic GWBs. We assume a network of four second-generation GW interferometers (Hanford, Livingston, Virgo, KAGRA). Crowder \textit{et al}. uses a Bayesian parameter estimation~\cite{Mandic:2012pj} under the assumption that $\Omega_\GW^{(I)} (f)$ follows a power law and derives 95\% confidence bounds on $v$. We do not show such bounds for the average chirp mass larger than $80 M_\odot$ as $\Omega_\GW^{(I)}$ cannot be described accurately by a simple power law (see the left panel of Fig.~\ref{fig:Omega-GW-I}). 
}
\label{fig:vaz}
\end{figure}

Figure~\ref{fig:vaz} compares bounds on the constant $v$ as a function of the average chirp mass with a network of four second-generation detectors using the analysis presented in this paper (red solid) and the Bayesian parameter estimation one in~\cite{Crowder:2012ik} (blue dashed). The former are obtained under an ideal situation that the magnitude of the GW spectrum is known a priori. The latter were obtained by simultaneously fitting for $v$ (or equivalently $\Pi$) and the overall magnitude of the GW spectrum $\Omega_{\alpha}$ with $\alpha=2/3$ and $f_\mrm{ref} = 100$Hz, though the correlation between these two parameters seems to be very small (see the bottom left panel of Fig.~2 in~\cite{Crowder:2012ik}). We do not present bounds from~\cite{Crowder:2012ik} for the average chirp mass larger than $80 M_\odot$ as $\Omega_\GW^{(I)}$ deviates significantly from a single power law.
Observe the qualitative agreement between the two analyses. For example, for the average chirp mass of $28M_\odot$ with which the single-power law is an excellent approximation for $\Omega_\GW^{(I)}$, the difference between the two analyses only give a $\sim 30$\% difference for the bound on $v$. Such an agreement justifies the validity of the analysis presented in this paper.

\section{Discussions}  
\label{sec:conclusion}

Let us now discuss possible avenues towards future extension of this work. One important direction is to derive bounds on CS gravity from solar system and binary pulsar observations without imposing $\ddot \vartheta = 0$. One can then compare such new bounds with the GW bounds presented in this paper and see if the former are complementary to the latter. It is likely that these experiments and observations are not as sensitive to $\ddot \vartheta$ as GW observations, but this point needs to be confirmed explicitly. 

We next discuss extensions applicable to both GWs from individual sources and GWBs. One obvious extension is to carry out more sophisticated analyses such as Bayesian parameter estimation studies for both GWs from individual sources~\cite{Cornish:2007if,Littenberg:2009bm} and stochastic GWBs~\cite{Crowder:2012ik,Mandic:2012pj}. 
Another future direction includes accounting for parity violation corrections from the generation of GWs in addition to the propagation effect that we have considered in this paper, and study how the former change the results presented here. For example, a gravitational parity-violating model has been proposed in~\cite{Contaldi:2008yz}, in which the right-handed and left-handed circular polarization modes couple to a different gravitational constant $G_{R,L}$. Since the propagation effect does not depend on $G_{R,L}$, the parity-breaking generation mechanism of GWs becomes the dominant effect to probe such a model.

Let us next explain possible future directions for GWs from individual sources. In this paper, we have assumed that the GW signals detected for the LVC events are consistent with GR and have carried out a Fisher analysis, which does not take into account systematics due to uncertainties in source parameters such as BH masses and spins. One important future work will be reanalyzing the calculations using the actual data. By doing so, one can place an upper bound on the scalar field evolution parameters to a given confidence level, instead of finding the \emph{distribution} of the upper bound on such parameters presented in this paper, which is an artifact of not using the actual data.  It would be also interesting to see how the bounds improve with future \emph{multiband} GW astronomy~\cite{Sesana:2016ljz,Barausse:2016eii} by combining ground-based and space-based observations. For example, space-based detectors significantly increases the angular resolution of GW sources~\cite{Sesana:2016ljz}, which allows one to partially break the degeneracy between source position/orientation parameters and parity violation effects.

For GWB-related observations, one possible avenue is to study how the bounds presented in this paper are affected by choosing different BH merger history. For example, one can consider GWBs from Population III (zero metallicity) binary BHs~\cite{Inayoshi:2016hco} with a larger chirp mass and higher redshift distribution than Population I or II binary BHs. 
It would be also important to study how the results change if one relaxes the equal-mass and spinless assumptions adopted in this paper.

Finally this paper focused on GWs from binary BH coalescences but it would be interesting to repeat the analysis for those from binary neutron star mergers. For example, the recent multimessenger observation of GW170817~\cite{Abbott:2017oio} tells us a precise sky localization of the source and that the inclination should be relatively small. Such additional information would partially break degeneracies between the CS parameter and angular parameters as discussed in~\cite{Yunes:2010yf} and may give an interesting bound on the former. 
GW170817 also suggests that the stochastic GWB from neutron star binaries may be comparable to that of BH binaries~\cite{Abbott:2017xzg}. Thus, one needs to study how the results presented here may change if one includes the contribution from neutron star binary GWBs.

\acknowledgments

We thank
Nicol\' as Yunes
for fruitful discussions and checking some of the expressions in the manuscript.
We also thank Katie Chamberlain for sharing with us the data file for the Voyager noise curve and Naoki Seto and Takahiro Tanaka for helpful discussions.
K.Y. and H.Y. acknowledge support from Simons Foundation.
K.Y. also acknowledges support from NSF grant PHY-1305682.
H.Y. also acknowledges support from NSF grant
PHY-1607449. 
This research was supported in part by NSERC, CIFAR and by the Perimeter Institute for Theoretical Physics. Research at Perimeter Institute is supported by the Government of Canada through the Department of Innovation, Science and Economic Development Canada, and by the Province of Ontario through the Ministry of Research and Innovation.

\bibliography{master}
\end{document}